\documentclass[%
twocolumn,
superscriptaddress,
 amsmath,amssymb,
 aps,
prc,
floatfix,
]{revtex4-1}

\usepackage{amsmath}
\usepackage{amssymb}
\usepackage{amsfonts}

\usepackage{xcolor}

\usepackage{dsfont}

\usepackage{braket}
\usepackage{graphicx}
\usepackage{dcolumn}
\usepackage{bm}

\newcommand{\bfvec}[1]{\mbox{\boldmath $#1$}}

\begin{document}

\preprint{APS/123-QED}

\title{Description of $^7$Be and $^7$Li within the Gamow Shell Model}

\author{J.P. Linares Fernandez}
\affiliation{Grand Acc\'el\'erateur National d'Ions Lourds (GANIL), CEA/DSM - CNRS/IN2P3, BP 55027, F-14000 Caen, France}
\author{N. Michel}
\affiliation{AS Key Laboratory of High Precision Nuclear Spectroscopy, Institute of Modern Physics, Chinese Academy of Sciences, Lanzhou 730000, China}
\affiliation{School of Nuclear Science and Technology, University of Chinese Academy of Sciences, Beijing 100049, China}
\author{M. P{\l}oszajczak}
\affiliation{Grand Acc\'el\'erateur National d'Ions Lourds (GANIL), CEA/DSM - CNRS/IN2P3, BP 55027, F-14000 Caen, France}
\author{A. Mercenne}
\affiliation{Department of Physics and Astronomy, Louisiana State University, Baton Rouge, LA 70803, USA}

\date{\today}

\begin{abstract} 
  \noindent {\bf Background:} $^7$Li and $^7$Be play an important role in Big Bang nucleosynthesis and nuclear astrophysics. The $^3$H($^4$He,$\gamma$)$^7$Li radiative capture reaction is crucial for the determination of the primordial $^7$Li abundance. In nuclear astrophysics, lithium isotopes have attracted a great interest because of the puzzled abundance of $^6$Li and $^7$Li.  \\
  {\bf Purpose:} In this work we study spectra of $^7$Be, $^7$Li and elastic scattering cross sections $^4$He($^3$He, $^3$He), $^4$He($^3$H, $^3$H) within the Gamow shell model (GSM) in the coupled-channel formulation (GSM-CC). The evolution of channel amplitudes and spectroscopic factors in the vicinity of the channel threshold is studied for selected states. \\
  {\bf Methods:} GSM provides the open quantum system formulation of nuclear shell model. In the representation of GSM-CC, GSM provides the unified theory of nuclear structure and reactions which is suited for the study of resonances in $^7$Be, $^7$Li and elastic scattering cross-sections involving $^3$H and $^3$He projectiles.\\ 
  {\bf Results:} The GSM-CC in multi-mass partition formulation applied to a translationally invariant Hamiltonian with an effective finite-range two-body interaction reproduce well the spectra of $^7$Be, $^7$Li and elastic scattering reactions: $^4$He($^3$He, $^3$He), $^4$He($^3$H, $^3$H). Detailed analysis of the dependence of reaction channel amplitudes and spectroscopic factors on the distance from the particle decay threshold allowed to demonstrate the alignment of the wave function in the vicinity of the decay threshold. This analysis also  demonstrates the appearance of clustering in the GSM-CC wave function in the vicinity of the cluster decay threshold. 
  \\
  {\bf Conclusions:} We demonstrated that GSM formulated in the basis of reaction channels including both cluster and proton/neutron channels allows to describe both the spectra of nuclei with low-energy cluster thresholds and the low-energy elastic scattering reactions with proton, $^3$H, and $^3$He projectiles. Studying dependence of the reaction channel amplitude and spectroscopic factor in a many-body state on a distance from the threshold, we showed an evolution of the $^3$He, $^3$H clustering with increasing separation energy from the cluster decay threshold and demonstrated a mechanism of the alignment of many-body wave function with the decay threshold \cite{Okolowicz2012,Okolowicz2013}, i.e. the microscopic reorganization of the wave function in the vicinity of the cluster decay threshold which leads to the appearance of clustering in this state.

\end{abstract}

\pacs{Valid PACS appear here}
\maketitle

\section{Introduction}
\label{intro}
The properties of radioactive nuclei are the basis for understanding nuclear mechanisms involved in astrophysical processes. These nuclei are studied in various reaction processes, and their properties are strongly affected by couplings to many-body continuum of scattering and decay channels. Therefore, the unified theory of nuclear structure and reactions is essential for the comprehensive description of radioactive nuclei in which bound states, resonances and scattering many-body states are treated equally and within a single theoretical framework. A pioneering work in this direction was initiated with the continuum shell model~\cite{barz1978,Rotter1978,rotter_review_1991,*Okolowicz2003,bennaceur_2000,rotureau_2005,*rotureau_2006,volya_2005}. Later, the \textit{ab initio} description of structure and reactions of light nuclei has been developed within the no-core shell model coupled with the resonating-group method (NCSM-RGM) \cite{quaglioni_2008,quaglioni_2009} and the no-core shell model with continuum (NCSMC) \cite{baroni_2013a,*baroni_2013b}.
One can recall that the standard approach of reaction theory is to build few-body nuclear systems from uncorrelated cluster wave functions, with which cross sections are calculated by using coupling potentials directly fitted on experimental data, in the frame of the R-matrix theory for example \cite{Descouvemont_2010}.

An alternative approach to describe radioactive nuclei within a unifying framework has been proposed with the Gamow Shell Model (GSM) \cite{Michel2002,Michel2003,michel_review_2009,michel_book_2021}, which provides the open quantum system formulation of nuclear shell model.
 GSM offers the most general treatment of couplings between discrete and scattering states, as it makes use of Slater determinants defined in the Berggren ensemble of single-particle states~\cite{berggren_1968}. GSM is the nuclear structure model {\it par excellence}.
 For the description of scattering properties and nuclear reactions, it is convenient to formulate GSM in the representation of reaction channels (GSM-CC) \cite{jaganathen_2014}. The calculation of resonances using this Hamiltonian is performed in the Berggren basis, so that the Hamiltonian matrix in GSM-CC is complex symmetric. However, the cross sections are calculated by coupling the real-energy incoming partial waves to the target states generated by a Hermitian shell model Hamiltonian. Consequently, cross sections are calculated in a fully Hermitian framework, whereas complex energies for resonances arise because one diagonalizes the complex symmetric Hamiltonian matrix in Berggren basis representation. In fact, GSM-CC makes use of the RGM framework \cite{wildermuth_tang_book_1977}, where a basis of reaction channels is built by coupling target and projectile states. The fundamental difference in GSM-CC is that target and projectile states are GSM eigenstates \cite{michel_book_2021}. Thus, GSM-CC is a complex-energy extension of the standard RGM approach of Ref.\cite{wildermuth_tang_book_1977}. GSM-CC approach has been applied to the description of proton elastic scattering \cite{jaganathen_2014}, deuteron elastic scattering \cite{mercenne_2019}, neutron transfer in $(d,p)$ reaction \cite{Mercenne2023}, and proton/neutron radiative capture reactions \cite{Fossez15,dong17,dong22}.

In this work, we study low-energy spectra of $^7$Li and $^7$Be in the binary multi-mass partition GSM-CC approach. The lowest energy thresholds correspond to the emission of neutron, $^3$H and proton, $^3$He, 
in $^7$Li and $^7$Be, respectively. Hence, in $^7$Li we deal with $^6$Li + $n$ and $^4$He + $^3$H mass partitions, in $^7$Be we consider $^6$Li + $p$ and $^4$He + $^3$He partitions.

$^7$Li and $^7$Be play an important role in Big Bang nucleosynthesis (BBN) and nuclear astrophysics. The $^3$H($^4$He,$\gamma$)$^7$Li radiative capture reaction is crucial for the determination of the primordial $^7$Li abundance. In nuclear astrophysics, lithium isotopes have attracted a great interest because of the puzzled abundance of $^6$Li and $^7$Li. Whereas $^7$Li in hot, low-metallicity stars is supposed to come from the BBN, $^6$Li is believed to originate from the spallation and fusion reactions in the interstellar medium \cite{Jedamzik20}. Therefore, the abundance ratio of $^6$Li and $^7$Li could be considered as an effective time scale of the stellar evolution \cite{Vangioni99}.

The  $^3$He($^4$He,$\gamma$)$^7$Be radiative capture reaction is essential for determining the fraction of branches in the $pp$-chain resulting in $^7$Be and $^8$B neutrinos \cite{Adelberger98,Adelberger11}. Much interest has been devoted to the study of reactions which can produce $^7$Be in the stellar environment \cite{Adelberger11}, especially to the ${^6}$Li($p$,$\gamma$)${^7}$Be reaction which is crucial for the elimination of $^6$Li and the formation of $^7$Be. Recent experimental studies of this reaction suggested a possible resonant enhancement of the ${^6}$Li($p$,$\gamma$)${^7}$Be cross section near threshold \cite{He13} (see also discussion of this reaction in Ref.\cite{dong17}). 

The existence of clustering and its survival in the competition with nucleonic degrees of freedom can be conveniently studied in the basis of reaction channels comprising both cluster and proton/neutron channels. The evolution of the channel probability in the wave function with the distance from the threshold of a given reaction channel can teach us how specific nucleon-nucleon correlations appear in the wave function and what the favored conditions are for their formation. Complementary information on the evolution of nucleon-nucleon correlations is provided by the change of the spectroscopic factors for nucleons and nucleon clusters with the distance from the reaction channel threshold. 

In this work, we study the near-threshold behavior of the probability weights of various reaction channels and the spectroscopic factors in selected states of $^7$Li and $^7$Be in the multi-mass partition coupled-channel framework of the GSM-CC.
To benchmark GSM-CC for elastic scattering with heavier projectiles, we study the reactions: $^4$He($^3$H,$^3$H)$^4$He and $^4$He($^3$He,$^3$He)$^4$He which probe the cluster structures in $^7$Li and $^7$Be. 
The paper is organized as follows. In chapter \ref{Formalism} we give the basic elements of the GSM-CC formalism, while in chapter \ref{model_space} we present the specific points related to the Hamiltonian (Sec.~\ref{Hamiltonian}) and the model space (Sec.~\ref{Model space}). Chapter \ref{Results} contains results. Spectra and structure of wave functions in $^7$Li and $^7$Be are displayed in Sec.~\ref{spectra7}. The behavior of channel amplitudes and spectroscopic factors for selected mirror states of $^7$Li and $^7$Be in the vicinity of channel threshold are discussed in Sec.~\ref{nearthrsec}. Differential cross sections for the elastic scattering reactions with proton, $^3$H, and $^3$He projectiles are presented in Sec.~\ref{crosssec}. Finally, in chapter \ref{conclusions} we summarize the main conclusions of this work.

\section{Theoretical framework}
\label{Formalism}
{
In this section we will briefly outline the GSM-CC formalism for the channels constructed with different mass partitions. Detailed discussion of the GSM-CC in a single mass partition case can be found in Refs.\cite{jaganathen_2014,mercenne_2019,michel_book_2021}.
GSM-CC has already been defined for the case of one-nucleon projectile in Ref.\cite{jaganathen_2014} and in the case of deuteron projectile in Ref.\cite{mercenne_2019}. Here, we will mainly concentrate on the differences between one and many mass partition case of the GSM-CC and apply for the description of spectra of mirror nuclei $^7$Be, $^7$Li and the mirror elastic scattering reactions $^4$He($^3$He,$^3$He)$^4$He and $^4$He($^3$H,$^3$H)$^4$He. 

We work in the cluster orbital shell model formalism (COSM), where all space coordinates are defined with respect to the center-of-mass (c.m.)~of an inert core \cite{Suzuki1988,michel_book_2021}. For valence nucleons, one has $\mathbf{r} = \mathbf{r_{\rm lab}} - \mathbf{R_{\rm c.m.}^{\rm (core)}}$, where
$\mathbf{r_{\rm lab}}$ is the nucleon coordinate in the laboratory frame, $\mathbf{R_{\rm c.m.}^{\rm (core)}}$ is the core c.m.~coordinate in the laboratory frame, which then define $\mathbf{r}$ as the nucleon coordinate in the COSM frame. The fundamental advantage of COSM is that one deals with translationally invariant space coordinates, so that no spurious center of mass motion can occur \cite{Suzuki1988,michel_book_2021}.

Clearly, the use of a core+valence nucleon picture leads to smaller model space dimensions than in NCSM-RGM and NCSMC \cite{quaglioni_2008,quaglioni_2009,baroni_2013a,*baroni_2013b}. Consequently, GSM and GSM-CC are more convenient numerically as only a few valence nucleons usually have to be considered therein in target states. When phenomenological Hamiltonians are utilized in GSM and GSM-CC, this also allows to systematically study the effect of interaction parameters on observables, which would be very difficult both theoretically and practically with no-core approaches. Additionally, the use of core coordinates greatly simplifies antisymmetry requirements, as Slater determinants built from one-body states in the COSM frame are both antisymmetric and without spuriosity \cite{Suzuki1988,michel_book_2021}. This is in contrast with other models, where one typically uses Jacobi coordinates for that matter \cite{wildermuth_tang_book_1977}, with which antisymmetry is cumbersome to apply in practice.

The ${ A }$-body state of the system is decomposed into reaction channels defined as binary clusters:
\begin{equation}
  \ket{ { \Psi }_{ M }^{ J } } = \sum_{\rm  c } \int_{ 0 }^{ +\infty } \ket{{ \left( {\rm c} , r \right) }_{ M }^{ J } } \frac{ { u }_{\rm c }^{JM} (r) }{ r } { r }^{ 2 } ~ dr \; ,
  \label{scat_A_body_compound}
\end{equation}
where the radial amplitude ${ {u}_{\rm c }^{JM}(r) }$, describing the relative motion between the two clusters in a channel ${\rm c }$, is the solution to be determined for a given total angular momentum ${J}$ and its projection ${M}$. 
The different channels in the sum of Eq.~(\ref{scat_A_body_compound}) are orthogonalized independently of their mass partition, involving different numbers of neutrons and protons.
The integration variable ${ r }$ is the relative distance between the c.m.~of the cluster projectile and that of the inert core \cite{michel_book_2021}, and the binary-cluster channel states are defined as:
\begin{equation}
\ket{ \left( {\rm c} , r \right)} = \hat{ \mathcal{A}} [\ket{\Psi_{\rm  T}^{J_{ \rm T }}; N_T, Z_T} \otimes  \ket{r ~ L_{\rm c.m.} ~ J_{\text{\rm int}} ~ J_{\rm P}; n, z}]_{ M }^{J} \label{channel}
\end{equation} 
where the channel index ${\rm c}$ stands for different quantum numbers and mass partitions $\{ (N_T, Z_T, { J }_{ \rm T}) ; (n , z, { L_{\rm c.m.} } , J_{\text{\rm int}}, J_{\rm P}) \}$,  $N_T$ and $Z_T$ are the number of neutrons and protons of the target, and $n$ and $z$ are the number of neutrons and protons of a projectile, so that $N = N_T + n$ and $Z = Z_T + z$ are the total number of neutrons and protons in the combined system of a projectile and a target. ${\hat{ \mathcal{A}}}$ is the inter-cluster antisymmetrizer that acts among the nucleons pertaining to different clusters. 
The states $\ket{\Psi_{\rm  T}^{J_{ \rm T }}} $ and $\ket{r ~ L_{\rm c.m.} ~ J_{\text{\rm int}} ~ J_{\rm P}}$ are the target and projectile states in the channel $\ket{ \left( {\rm c} , r \right)}$ of Eq.~(\ref{channel}) with their associated total angular momentum ${ { J }_{ \rm T } }$ and ${ { J }_{ \rm P } }$, respectively.
The angular momentum couplings read $\mathbf{ J}_{\rm P } = \mathbf{ J}_{ \rm int } + \mathbf{L}_{\rm c.m.}$ and  ${ \mathbf{ J}_{\rm A} = \mathbf{J}_{\rm P} + \mathbf{ J}_{\rm T} } $. Quantum numbers of many-body projectiles are customarily denoted by $^{2J_{\rm int}+1}(L_{\rm c.m.})_{J_{\rm P}}$ in numerical applications. These angular quantum numbers will also be denoted by $\ell j$ when dealing with one-nucleon systems for clarity.

The Schr{\"o}dinger equation $H \ket{\Psi_{M}^{J}} = E \ket{\Psi_{M}^{J}}$ in the channel representation of the GSM takes the form of coupled-channel equations:
\begin{equation}
  \sum_{\rm c}\int_{0}^{\infty}  \!\!\! r^{ 2 } \left( H_{\rm cc' } (r , r') - E N_{\rm cc' } (r , r') \right) \frac{ { u }_{\rm c } (r) }{ r } = 0	\ ,
  \label{cc_cluster_eq}
\end{equation}
where ${ E }$ stands for the scattering energy of the ${ A }$-body system. To simplify reading, we have dropped the total angular momentum labels ${ J }$ and ${ M }$, but one should keep in mind that Eq.~(\ref{cc_cluster_eq}) is solved for fixed values of ${J}$ and ${ M }$.
The kernels in Eq.~(\ref{cc_cluster_eq}) are defined as:
\begin{align}
  & H_{\rm cc' } (r,r') = \bra{ ({\rm c},r) } \hat{ H } \ket{({\rm c'},r') } \label{h_cc_compound} \\
  & N_{\rm cc' } (r,r') = \braket{ ({\rm c},r) | ({\rm c'},r') } \label{n_cc_compound}
\end{align}

As the nucleons in the target and those of the projectile interact via a short-range interaction in our model, it is convenient to express the Hamiltonian $\hat{ H }$ as:
\begin{equation}
  \hat{ H } = \hat{ H }_{ \rm T } + \hat{ H }_{ \rm P } + \hat{ H }_{ \rm TP }		
  \label{new_hamiltonian}
\end{equation}
where ${ \hat{ H }_{ \rm T } }$ and ${ \hat{ H }_{ \rm P } }$ are the Hamiltonians of the target and projectile, respectively.
In particular, ${ { \hat{ H } }_{ \rm T } }$ is the intrinsic Hamiltonian of the target, and its eigenvectors are ${ \ket{ { \Psi }_{ \rm T }^{ { J }_{ \rm T } } } }$ with eigenvalues ${ { E }_{ \rm T }^{ { J }_{ \rm T } } }$. ${ \hat{ H } }$ is considered to be the standard GSM Hamiltonian.
The projectile Hamiltonian ${ { \hat{ H } }_{ \rm P } }$ can be decomposed as follows: ${ { \hat{ H } }_{ \rm P } = { \hat{ H } }_{ \text{\rm int} } + { \hat{ H } }_{ \text{c.m.} } }$.
${ { \hat{ H } }_{\rm  int } }$ describes the intrinsic properties of the projectile and $\ket{J_{\rm int}}$ is its eigenvector with an eigenvalue ${ { E }_{ \rm int }^{ { J }_{ \rm int } } }$.
${ { \hat{ H } }_{ \rm c.m. } }$ describes movement of the projectile center-of-mass, defined in a single channel $c$ as:
 \begin{equation}
   { \hat{ H } }_{\rm c.m. } = \frac{ { \hbar }^{ 2 }}{ 2 \Tilde{m}_{\rm P}} \left( -\frac{ { d }^{ 2 }}{ d r^{ 2 }} + \frac{L(L + 1)}{r^2} \right)  + { U_{\rm c.m.}^{ L }} (r) \ ,
  \label{HCM_definition}
\end{equation} 
where $L=L_{\rm c.m.}$ is the c.m.~orbital angular momentum, ${ {\Tilde{m} }_{ \rm P } }$ in this equation is the reduced mass of the projectile and ${ { U }_{\rm c.m. }^{ L }(r) }$ is the basis-generating Woods-Saxon (WS) potential for nucleon projectile, while it is the weighted sum of proton and neutron basis-generating WS potentials for the multi-nucleon projectile wave functions \cite{mercenne_2019,michel_book_2021}:
\begin{eqnarray}
\!\!\!\!\!\!\!\!\!\!\!\! U^{ L }_{{\rm c.m.}, {\rm C}} (r) &=& z~U^{ L }_{\rm p, C} (r) + n~U^{ L }_{\rm n, C} (r)  
\label{UCM_central} \\
\!\!\!\!\!\!\!\!\!\!\!\! U^{ L }_{\rm c.m., s.o. } (r) &=& \frac{z}{n+z}~U^{ L }_{\rm p, s.o.} (r) + \frac{n}{n+z}~U^{ L }_{\rm n, s.o.} (r) 
 \label{UCM_so_average} \ ,
\end{eqnarray}
where $z$ and $n$ are the number of protons and neutrons of the cluster, respectively, while $U^{ L }_{\rm p, C} (r)$, $U^{ L }_{\rm p, s.o.} (r)$ and $U^{ L }_{\rm n, C} (r)$, $U^{ L }_{\rm n, s.o.} (r)$ are the WS basis-generating central and spin-orbit potentials for proton and neutron, respectively. 

One uses fractional masses in front of neutron and proton spin-orbit potentials in order to form an average spin-orbit potential in Eq.~(\ref{UCM_so_average}). Indeed, from this averaging procedure, one can separate the radial spin-orbit dependence from the sum over the $\bfvec{\ell} \cdot \bfvec{s}$ terms of the nucleons of the cluster projectile. The spin of the projectile then appears explicitly : $\sum \bfvec{\ell} \cdot \bfvec{s} \simeq (1/(n+z)) ~ \mathbf{L}_{\rm c.m.}  \cdot \sum \bfvec{s} \simeq (\mathbf{L}_{\rm c.m.} \cdot \mathbf{J_{\rm int}})/(n+z)$. An additional factor $1/(n+z)$ arises because $\ell \simeq L_{\rm c.m.}/(n+z)$ \cite{mercenne_2019,michel_book_2021}.

The potential $U^{ L }_{\rm c.m.}(r)$ of Eq.~(\ref{HCM_definition}) then reads:
\begin{eqnarray}
U^{ L }_{\rm c.m.} (r) &=& U^{ L }_{\rm c.m., C} (r) \nonumber \\
&+& \frac{1}{n+z} ~ { U^{ L }_{\rm c.m., s.o. }} (r) ~ (\mathbf{L}_{\rm c.m.} \cdot \mathbf{J_{\rm int}}) \ .
\label{UCM}
\end{eqnarray}

In order to calculate the kernels $H_{\rm cc' } (r,r')$ and $N_{ \rm cc' } (r,r')$ (Eqs.~(\ref{h_cc_compound}) and (\ref{n_cc_compound})), one expands ${ \ket{ (c,r) } }$ onto a one-body Berggren basis:
\begin{equation}
  \ket{ ({\rm c},r) } = \sum_{N_{\rm c.m.}} \frac{ { u }_{ N_{\rm c.m.}} (r) }{ r } \ket{ ({\rm c},N_{\rm c.m.}) } \ ,
  \label{expansion_channel_n}
\end{equation}
where $N_{\rm c.m.}$ refers to the projectile c.m.~shell number in the Berggren basis generated by diagonalizing ${ { \hat{ H } }_{\rm c.m. } }$ (see Eqs.~(\ref{HCM_definition},\ref{UCM})), i.e.~${ { \hat{ H } }_{\rm c.m. } \ket{ N_{\rm c.m.} ~ L_{\rm c.m.} } = { E }_{\rm  c.m. } \ket{ N_{\rm c.m.} ~ L_{\rm c.m.} } }$,
and where $${ \ket{ ({\rm c},N_{\rm c.m.}) } = \hat{ \mathcal{A}} \ket{ \{ \ket{\Psi_{ \rm T }^ {J_{ \rm T } }} \otimes \ket{N_{\rm c.m.} ~ L_{\rm c.m.} ~ J_{\text{int}} ~ J_{\rm P}} \}_{ M }^{J}} } \ .$$ For simplicity, the channel dependence has been omitted in the notation of $u_{N_{\rm c.m.}}(r)$.

Consequently, one can expand Eqs.~(\ref{h_cc_compound},\ref{n_cc_compound}) onto the basis of ${ \ket{ ({\rm c},N_{\rm c.m.}) } }$ using Eq.~(\ref{expansion_channel_n}) and derive the following expression for the Hamiltonian and norm kernels:
\begin{eqnarray}
  H_{\rm cc' } (r, r') &=& \left( { \hat{ H } }_{\rm c.m. } + E_{\rm T }^{ { J }_{\rm T } } + {E}_{\rm P}^{ { J }_{ \text{int} } } \right) \frac{ \delta (r - r') }{ r r' } { \delta }_{\rm cc' } \nonumber \\
  &+&  { \tilde{ V }}_{\rm cc' } (r , r')
  \label{hamiltonian_matrix_elmts} \\
  N_{\rm cc' } (r,r') &=&  \frac{ \delta (r - r') }{ r r' } { \delta }_{\rm cc' }  + \Delta N_{\rm cc' } (r,r') \label{norm_matrix_elmts}
\end{eqnarray}
where $\tilde{ V }_{\rm cc' }(r,r')$ includes the remaining short-range potential terms of the Hamiltonian kernels and is the inter-cluster potential and $\Delta N_{\rm cc' } (r,r')$ is a finite-range operator as well. $H_{\rm cc' } (r, r')$ reduces to its diagonal part at large distance as $\tilde{ V }_{\rm cc' }(r,r')$ vanishes identically at a large but finite radius outside the target. Hence, nucleon transfer, which is induced by $\tilde{ V }_{\rm cc' }(r,r')$, and consequently ${ { \hat{ H } }_{\rm TP } }$, can only occur in the vicinity of the target and not in the asymptotic region.

The determination of ${ { \tilde{ V } }_{\rm cc' }(r,r') }$ involves the calculation of the matrix elements of ${ { \hat{ H } }_{\rm TP } }$, which contain a shell model Hamiltonian.
In order to compute ${ { \hat{ H } }_{\rm TP } }$ one has to expand each ${ \ket{ ({\rm c},N_{\rm c.m.}) } }$ onto a basis of Slater determinants built upon single-particle (s.p.) states of the Berggren ensemble.
In practice, the intrinsic target and projectile states, ${ \ket{ { \Psi }_{\rm T }^{ { J }_{\rm T } } } }$ and ${ \ket{ { J }_{\rm int } } }$ respectively, are already calculated with that basis, as ${ { \hat{ H } }_{\rm T } }$ and ${ { \hat{ H } }_{ \text{int} } }$ are solved using the GSM. Consequently, the nuclear structure of target and projectile states is more realistic than in the R-matrix theory, for example, where nuclei are typically built from a few uncorrelated clusters \cite{Descouvemont_2010}.
Note that in general, as we deal with very light projectiles, ${ { \hat{ H } }_{ \text{int} } }$ is solved within a no-core framework, and this will be the case in the present study.
The remaining task consists in expanding ${ \ket{ N_{\rm c.m.} ~ L_{\rm c.m.} ~ { J }_{\rm int } } }$ in a basis of Slater determinants. In GSM-CC, one applies for this a center-of-mass excitation raising operator onto ${ \ket{ { J }_{\rm int } } }$. 
More details can be found in Refs.\cite{mercenne_2019,michel_book_2021}.

The many-body matrix elements of the norm kernel Eq.~(\ref{n_cc_compound}) are calculated using the Slater determinant expansion of the cluster wave functions ${ \ket{ ({\rm c},N_{\rm c.m.}) } }$.
Note that the antisymmetry of channels, enforced by the antisymmetrizer in Eq.~(\ref{channel}), is exactly taken into account through the expansion of many-body targets and projectiles with Slater determinants.

The treatment of the non-orthogonality of channels is the same as in the one-nucleon projectile case \cite{jaganathen_2014}.
Channels are indeed not orthogonal in general because of the antisymmetrizer in Eq.~(\ref{channel}). In order to deal with orthogonalized channels, $N_{\rm cc' } (r , r')$ (see Eq.~(\ref{n_cc_compound})) is diagonalized, which generates orthogonal channels. These channels are linear combinations of the initial channels $\ket{ \left( {\rm c} , r \right)}$ (see Eq.~\ref{channel}). The coupled-channel equations of Eq.~\ref{cc_cluster_eq} then become :
\begin{eqnarray}
  \sum_{\rm c}\int_{0}^{\infty}  \!\!\! && r^{ 2 } \left( \widetilde{H}_{\rm cc' } (r , r') - E \delta_{cc'} \frac{\delta(r-r')}{rr'} \right) \nonumber \\ 
  &\times& \frac{ { w }_{\rm c } (r) }{ r } = 0	\ ,
  \label{cc_cluster_eq_orthog}
\end{eqnarray}
where $\tilde{H}_{\rm cc' } (r , r')$ contains $H_{\rm cc' } (r , r')$ and terms induced by the orthogonalization of channels, and the ${ w }_{\rm c } (r)$ functions are the orthogonalized channel wave functions \cite{jaganathen_2014}.

Once the kernels are computed, the coupled-channel equations of Eqs.~(\ref{cc_cluster_eq},\ref{cc_cluster_eq_orthog}) can be solved using a numerical method based on a Berggren basis expansion of the Green's function ${ { (H - E) }^{ -1 } }$, that takes advantage of GSM complex energies. Details of this method can be found in Refs. \cite{mercenne_2019,michel_book_2021}.

\section{Model space and Hamiltonian}
\label{model_space}
{
GSM in the Slater determinant representation will be used to optimize the effective interaction. In the following, the GSM-CC with cluster projectiles will be applied to the spectra of $^7$Be, $^7$Li, and elastic scattering reactions $^4$He($^3$He,$^3$He) and $^4$He($^3$H,$^3$H). 
}

\subsection{Effective Hamiltonian}
\label{Hamiltonian}

The effective Hamiltonian is optimized using the GSM in Berggren basis. The model is formulated in the relative variables of COSM that allow to eliminate spurious c.m.~excitations (see Sect.\ref{Formalism}).
We use $^4$He as an inert core with two, three or four valence nucleons to describe $^6$Li, $^7$Be and $^7$Li wave functions, respectively. 

The Hamiltonian consists of the one-body part and the nucleon-nucleon interaction of FHT type~\cite{1979Furutani,1980Furutani} supplemented by the Coulomb term:
\begin{equation}
V_{\rm FHT} = V_{\rm c} + V_{\rm LS} + V_{\rm T} + V_{\rm Coul} \label{V_FHT}
\end{equation}
where $V_{\rm c}$ , $V_{\rm LS}$, $V_{\rm T}$ represent its central, spin-orbit and tensor part, respectively. The two-body Coulomb potential 
$V_{\rm Coul}(r) = e^2/r$ between valence protons is treated exactly at GSM-CC level by incorporating its long-range part into the basis potential (see Ref.\cite{PRC_isospin_mixing} for a detailed description of the method). 
 
The $^4$He core is mimicked by a one-body potential of the WS type, with a spin-orbit term, and a Coulomb field (see Table \ref{WS_parameters}). 
The WS potential depth $V_0$, the spin-orbit strength $V_{\ell s}$, the radius $R_0$, and the diffuseness $a$ are the four parameters that enter the optimization carried out independently for protons and neutrons. The Coulomb potential was kept fixed and equal to the potential generated by a spherical Gaussian charge distribution: $U_{\rm Coul}(r)=2e^2{\rm erf}(r/{\tilde R}_{\rm ch})/r$~\cite{Sai77},
where ${\tilde R}_{\rm ch} = 4R_{\rm ch} /(3\sqrt{\pi} )$ and ${\rm erf}(r/{\tilde R}_{\rm ch})$ is the error function in $r/{\tilde R}_{\rm ch}$. The previous value for ${\tilde R}_{\rm ch}$ allows $U_{\rm Coul}(r)$ to resemble the Coulomb potential generated by a uniformly charged distribution of radius $R_{\rm ch}$.

The different components $V_{\rm FHT}$ in Eq.~(\ref{V_FHT}) read~\cite{PRC_FHT_Yannen}:
\begin{eqnarray}
\!\!\!\!\!V_{\rm c}(r) &=&  \sum_{n=1}^3  \, V_{\rm c}^n   \,    \left( W_{\rm c}^n + B_{\rm c}^n P_{\sigma} - H_{\rm c}^n  P_{\tau} \right. \nonumber \\
&& \qquad\qquad\quad \;\:\;\left. -~ M_{\rm c}^n P_{\sigma}P_{\tau} \right)\: e^{-\beta_{\rm c}^n r^2}  \label{FHT1} \\
\!\!\!\!\!V_{\rm LS}(r) &=& \mathbf{L}_{\rm rel}\cdot \mathbf{S} \sum_{n=1}^2 \, V_{\rm LS}^n  \, \left( W_{\rm LS}^n - H_{\rm LS}^n  P_{\tau}  \right) \, e^{-\beta_{\rm LS}^n r^2}   \label{FHT2} \\
\!\!\!\!\!V_{\rm T}(r) &=&  S_{ij}\:\sum_{n=1}^3 \,  V_{\rm T}^n \, \left( W_{\rm T}^n  - H_{\rm T}^n  P_{\tau} \right) \, r^2 e^{-\beta_{\rm T}^n r^2},   \label{FHT3}
\end{eqnarray}
where $r$ is the distance between the nucleons $i$ and $j$, 
$\mathbf{L}_{\rm rel}$ and $\mathbf{S}$ are the relative orbital angular momentum and spin of the two-nucleon system, respectively,
$S_{ij}=3 (\bfvec{\sigma}_i \cdot \mathbf{\hat{r}}) (\bfvec{\sigma}_j \cdot \mathbf{\hat{r}})   - \bfvec{\sigma}_i \cdot \bfvec{\sigma}_j$ is the tensor operator, 
$P_{\sigma}$ and  $P_{\tau}$ are spin and isospin exchange operators, respectively,
$V_{\rm c}^n$, $n \in \{1,2,3\}$ and $V_{\rm LS}^n$, $V_{\rm T}^n$, $n \in \{1,2\}$ are parameters fitting the central, spin-orbit and tensor parts, respectively, while other parameters are fixed~\cite{1979Furutani}.
Following Ref.\cite{PRC_FHT_Yannen}, we rewrite $V_{\rm FHT}$ in terms of its spin and isospin dependence:
\begin{eqnarray}
V_{\rm c}(r) &=&  V_{\rm c}^{11} \, f_{\rm c}^{11}(r)  \Pi_{11} \: + \:  V_{\rm c}^{10} \, f_{\rm c}^{10}(r) \Pi_{10} \nonumber  \\
&+& V_{\rm c}^{00} \, f_{\rm c}^{00}(r) \Pi_{00} \: + \:  V_{\rm c}^{01} \, f_{\rm c}^{01}(r) \Pi_{01},  \label{inter1}  \\
V_{\rm LS}(r) &=& (\mathbf{L}_{\rm rel}\cdot \mathbf{S}) \,V_{\rm LS}^{11} \, f_{\rm LS}^{11}(r) \Pi_{11}, \label{inter2}  \\
V_{\rm T}(r) &=& S_{ij} \left[V_{\rm T}^{11}f_{\rm T}^{11}(r) \Pi_{11} + V_{\rm T}^{10} f_{\rm T}^{10}(r) \Pi_{10}\right],\label{inter3}
\end{eqnarray}
where $\Pi_{\rm ST}$ are projectors on spin and isospin~\cite{1974DeShalit,RingSchuck} and $f_{\rm c}^{\rm ST}(r)$, $f_{\rm LS}^{\rm ST}(r)$ and  $f_{\rm T}^{\rm ST}(r)$ functions are straightforward to evaluate from Eqs.~(\ref{FHT1},\ref{FHT2},\ref{FHT3}). 

\begin{table}[htb]
\caption{\label{WS_parameters} 
Parameters of the one-body potential for protons and neutrons optimized to describe spectra of $^6$Li, $^7$Be, $^7$Li in GSM.  From top to bottom: central potential depth, spin-orbit potential depth, radius, diffuseness and charge radius.}
\begin{ruledtabular}
\begin{tabular}{lcc}
Parameter & Neutrons & Protons \\ \hline \\
$V_0$  (MeV) &  52.9   & 52.2\\
$V_{\ell s}$ (MeV\, fm$^2$) &  3.39  &3.77 \\
$R_0$ (fm) &  2.0 &  2.0 \\
$a$ (fm)  & 0.65 &  0.65 \\
$R_{\rm ch}$ (fm)  &--  & 2.5
\end{tabular}
\end{ruledtabular}
\end{table}
The matrix elements of the Hamiltonian are calculated in the model space consisting of all proton and neutron harmonic oscillator (HO) states having $\ell \leq 3$ and $n \leq 4$. The use of Berggren basis at this level is not necessary as the Slater determinants used therein only generate the GSM-CC Hamiltonian interaction ${ \tilde{ V }}_{\rm cc' } (r , r')$ (see Eq.~(\ref{hamiltonian_matrix_elmts})), which is finite-ranged.
Note that the use of HO states does not hamper the asymptotes of the loosely bound and resonance GSM-CC states of $^7$Be and $^7$Li. Indeed, HO states are used only to generate the finite range part of the GSM-CC Hamiltonian, whereas the GSM-CC eigenstates of $^7$Be, $^7$Li and $^8$Be are expanded with Berggren basis states. Consequently, the density of $^7$Be, $^7$Li GSM-CC eigenstates slowly decreases or increases exponentially in modulus, respectively, independently of the Gaussian fall-off of HO states. Convergence for Hamiltonian representation is typically obtained with 5-10 HO states per partial wave~\cite{PRC_isospin_mixing}.

\begin{table}[htb]
\caption{\label{FHT_parameters} 
Parameters of the FHT interaction used in this study are compared with the original FHT parameters with their statistical uncertainties reported in~\cite{PRC_FHT_Yannen} for $p$-shell nuclei. 
 }
\begin{ruledtabular}
\begin{tabular}{lcc}
Parameter   & FHT~\cite{PRC_FHT_Yannen} & FHT (this work)     \\ \hline \\ 
$V_{\rm c}^{11}$   & -3.2 (220) & -13.8 (232)   \\
$V_{\rm c}^{10}$   & -5.1 (10) & -5.38 (0.24)  \\
$V_{\rm c}^{00}$   & -21.3 (66) & -31.5 (145)   \\
$V_{\rm c}^{01}$   & -5.6 (5) & -5.3 (0.33)    \\
$V_{\rm LS}^{11}$ & -540 (1240) & -249.2 (2.7)   \\
$V_{\rm T}^{11}$  & -12.1 (795) & -11.1 (29)   \\
$V_{\rm T}^{10}$  & -14.2 (71) & -0.05 (4.7)    
\end{tabular}
\end{ruledtabular}
\end{table}
The parameters of the optimized one-body potential which imitate the effect of $^4$He core are shown in Table \ref{WS_parameters}.
The statistical properties of the FHT interaction parameters for $p$-shell nuclei have been analyzed in Ref.\cite{PRC_FHT_Yannen}. It has been noticed that they bear a sizable statistical error. Consequently, one can modify the FHT interaction parameters within the bounds of calculated statistical errors without in principle changing the interaction. 
Table \ref{FHT_parameters} compares parameters of the FHT interaction optimized in this work with those given in Ref.\cite{PRC_FHT_Yannen}. The optimization of the FHT interaction in GSM has been performed in a model space spanned by $spdf$ partial waves. 
One can see that the parameters of the interaction obtained in the present optimization agree with those of Ref.\cite{PRC_FHT_Yannen} within the statistical errors.

\subsection{Model space in GSM-CC calculation}
\label{Model space}

GSM calculations of $^6$Li target are performed in the approximation of $^4$He core, whereby $0s_{1/2}$ HO shells are fully occupied, and valence particles in two main resonant-like HO shells $0p_{3/2}$, $0p_{1/2}$, and several scattering-like subdominant HO shells in $\{s_{1/2}\}, \{p_{3/2}\}$, $\{p_{1/2}\}$ $\{d_{5/2}\}$, $\{d_{3/2}\}$, $\{f_{7/2}\}$, $\{f_{5/2}\}$ partial waves, verifying $n \in [1:4]$ in $sp$ waves and $n \in [0:4]$ in other partial waves. This small number of HO states per partial wave has been seen to satisfactorily approximate the non-resonant continuum. To reduce the size of the GSM matrix, the basis of Slater determinants is truncated by limiting the excitation energy to 8$\hbar\omega$ and restricting to 2 the number of nucleons in the continuum-like states. 
 \par

Antisymmetric eigenstates of the GSM-CC are expanded in the basis of channels: $[{^4}{\rm He}(0^+_1)\otimes{^3}{\rm He}(L_{\rm c.m.}~J_{\rm int}~J_{\rm P})]^{J^{\pi}}$ and $[{^6}{\rm Li}((J^\pi)_T)\otimes{p}(\ell j) ]^{J^{\pi}}$ for $^7$Be, and  $[{^4}{\rm He}(0^+_1)\otimes{^3}{\rm H}(L_{\rm c.m.}~J_{\rm int}~J_{\rm P})]^{J^{\pi}}$ and $[{^6}{\rm Li}((J^\pi)_T)\otimes{n}(\ell j)]^{J^{\pi}}$ for $^7$Li, where $^6$Li can be in its ground state or in low-lying excited states.

The internal structure of $^3$He and $^3$H projectiles in the channels $[{^4}{\rm He}(0^+_1)\otimes{^3}{\rm He}(L_{\rm c.m.}~J_{\rm int}~J_{\rm P})]^{J^{\pi}}$ and $[{^4}{\rm He}(0^+_1)\otimes{^3}{\rm H}(L_{\rm c.m.}~J_{\rm int}~J_{\rm P})]^{J^{\pi}}$ is calculated using the N$^3$LO interaction~\cite{PRC_N3LO} without the three-body contribution, fitted on phase shifts properties of proton-neutron elastic scattering reactions. The N$^3$LO realistic interaction is diagonalized in six HO shells to generate the intrinsic states of $^3$He  and $^3$H. The oscillator length in this calculation is $b=1.65$ fm. For this value of the oscillator length, the 
ground state energy of $^3$He equals -6.35 MeV, whereas the experimental value is -7.71 MeV. For $^3$H, the ground  state energy is -7.14 MeV as compared to -8.48 MeV experimentally. In the coupled-channel equations of GSM-CC, we use the experimental binding energies of $^3$He, $^3$H to assure correct thresholds $^4$He + $^3$He, $^4$He + $^3$H.

The relative motion of the $^3$He ($^3$H) cluster c.m.~and the $^4$He target is calculated in the Berggren basis
generated  by proton and neutron Woods-Saxon potentials.
Different $L_{\rm c.m.}=0,1,2,3$ partial waves bear 3, 3, 2, 2 pole states, respectively, which are included along with the respective contours. The contours consist of three segments, defined by the origin of the $K_{\rm c.m.}$ complex plane and the complex points $K_{\rm c.m.}$: 0.2-i0.1 fm$^{-1}$, 1.0-i0.1 fm$^{-1}$ and 2 fm$^{-1}$. Each segment is discretized with 15 points, so that each contour possesses 45 points.
All unbound pole states lie below the Berggren basis contours, so that they do not belong to the considered Berggren bases.
For the intrinsic $^3$He and $^3$H wave functions 
we consider only the most important one bearing $\left( J^{\pi} \right)_{\text{int}} = 1/2^+$. Since the c.m.~parts of $^3$He and $^3$H projectiles bear $L_{\rm c.m.} \leq 3$, therefore the total angular momentum of $^3$He and $^3$H projectiles satisfies $J_{\rm P} \leq 7/2$.

The use of two different interactions to deal with the structure of $^7$Be ($^7$Li) and the elastic scattering of $^3$He ($^3$H) on $\alpha$-particle is necessary as we have two different pictures in our model. Before and after the reaction occurs, $^3$He ($^3$H) is far from the target and its properties as a cluster projectile are prominent, whereas during the reaction the properties of composite systems $^7$Be ($^7$Li) are decisive. As the FHT interaction is defined from $^6$Li, $^7$Be, $^7$Li properties, it cannot grasp the structure of $^3$He ($^3$H) at large distances.
Conversely, the N$^3$LO interaction cannot be used in a core and valence particles approximation. Added to that, the N$^3$LO interaction generating $^3$He ($^3$H) projectiles makes use of laboratory coordinates, which are directly replaced by COSM coordinates when building channels (see Eq.~(\ref{channel})).  
While it would be possible, in principle, to apply the transformation from laboratory to COSM coordinates from a numerical point of view, this procedure would be very cumbersome, with its effect also much smaller than the other theoretical assumptions present in our model (presence of an inert core, use of an effective FHT interaction fitted from experimental data, restricted number of channels, truncated model spaces, etc). Moreover, as the N$^3$LO interaction enters only the $^3$He ($^3$H) projectile basis construction, it is not explicitly present in the Hamiltonian, but just insures that the projectile $^3$He ($^3$H) has both the correct wave function (binding energy) and asymptotic behavior. This also implies that the use of both laboratory and COSM coordinates is consistent therein, as they coincide asymptotically. Indeed, cross sections are always calculated in the asymptotic region, whereby $\mathbf{r_{\rm lab}} - \mathbf{R_{\rm c.m.}^{\rm (core)}} \simeq \mathbf{r_{\rm lab}}$ as $r_{\rm lab} \rightarrow +\infty$ therein and $\braket{R_{\rm c.m.}^{\rm (core)}}$ equals a few fm. As a consequence, the use of both realistic interaction for projectiles and effective Hamiltonian for composites induces no problem in the GSM-CC framework.

\section{Results}
\label{Results}

\subsection{Spectroscopy of $^7$Be and $^7$Li}
\label{spectra7}
The lowest particle emission thresholds in $^7$Be and $^7$Li are $^4$He + $^3$He and  $^4$He + $^3$H. Therefore, the description of low-energy states in $^7$Be  and $^7$Li requires the inclusion of the coupling to $^3$He and $^3$H continua, respectively. In this section, we shall discuss the spectra of mirror nuclei $^7$Be and $^7$Li in the channel basis comprising $[{^4}{\rm He}(0^+_1)\otimes{^3}{\rm He}(L_{\rm c.m.}~J_{\rm int}~J_{\rm P})]^{J^{\pi}}$, $[{^6}{\rm Li}((J^\pi)_T)\otimes{p}(\ell j )]^{J^{\pi}}$ in $^7$Be, and 
$[{^4}{\rm He}(0^+_1)\otimes{^3}{\rm H}(L_{\rm c.m.}~J_{\rm int}~J_{\rm P})]^{J^{\pi}}$, $[{^6}{\rm Li}((J^\pi)_T)\otimes{n}(\ell j)]^{J^{\pi}}$ in $^7$Li. 

\begin{figure*}[htb]
\begin{center}
\includegraphics[width=18cm,angle=0]{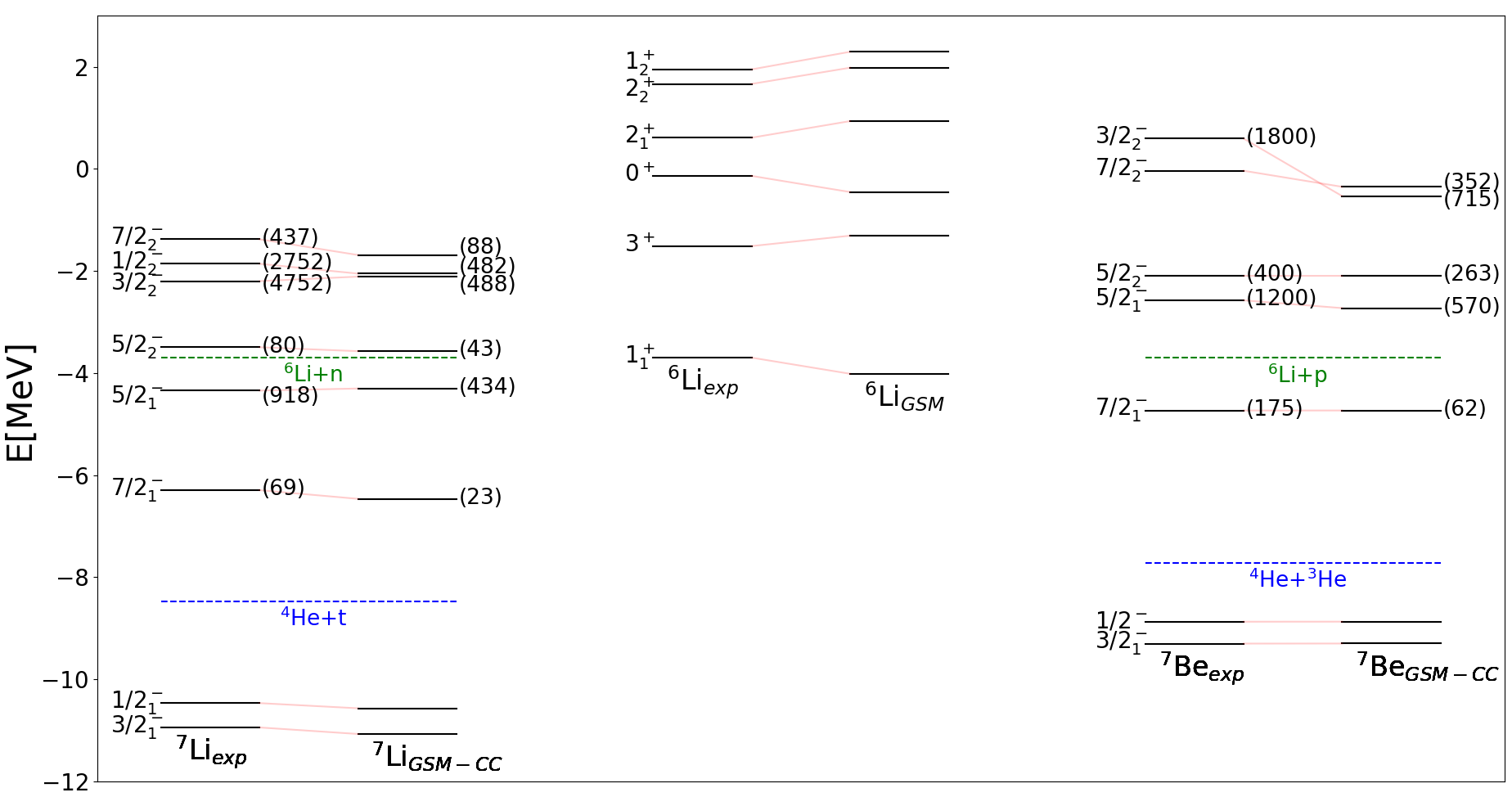}
\end{center}
\caption{(Color online) The calculated energy spectra  of $^{6,7}$Li and $^7$Be, are compared with experimental data \cite{ensdf}. $^{7}$Li and $^7$Be are calculated in GSM-CC using the channel basis with two mass partitions: $^{4}{\rm He} + {^{3}{\rm He}}$, $^6{\rm Li} + {p}$ and $^{4}{\rm He} + {^{3}{\rm H}}$, $^6{\rm Li} + {n}$, respectively. The spectrum of $^6$Li is calculated in GSM and wave functions are expanded in the HO basis. Numbers in the brackets indicate the resonance width in keV. 
}
\label{spec_7Li7Be}
\end{figure*}

Figure \ref{spec_7Li7Be} shows the GSM spectrum of $^6$Li and the GSM-CC spectrum of $^{7}$Li and $^7$Be. Wave functions of $^6$Li states shown in this figure are used to build channel states in $^7$Li and $^7$Be. The GSM-CC calculation for $^7$Be and $^7$Li are performed in the channel basis consisting of the two mass partitions: $^6$Li + $p$, $^3$He + $^4$He and $^6$Li + $n$, $^3$H + $^4$He, respectively. The interaction matrix elements entering the microscopic channel-channel coupling potentials involving nucleon-projectile channels have been rescaled by the tiny multiplicative corrective factors $c(J^\pi)$ for $3/2^-_1$, $1/2^-_1$, $7/2^-_1$, and $5/2^-_2$ states to correct for missing channels in the model space. These scaling factors are: $c(3/2^-)=1.0092$, $c(1/2^-)=1.0156$, $c(7/2^-)=1.0173$, $c(5/2^-)=0.9955$. 

The agreement with experimental data for resonance energies is excellent. However, the calculated widths are smaller than found experimentally. One may notice a change of the order of higher lying levels $1/2^-_2$, $3/2^-_2$, $7/2^-_2$ between $^7$Be and $^7$Li due to different threshold energies and Coulomb energies. All energies of the states are given relatively to the energy of $^4$He core. 
Experimental and calculated particle emission thresholds: $^4$He + $^3$He and $^6$Li + $p$ in $^7$Be and $^4$He + $^3$H and $^6$Li + $n$ in $^7$Li, coincide within the line thickness in the figure. Remaining tiny differences are corrected in the coupled-channel equations so that the threshold energies correspond exactly to the  experimental values. 

Channels in $^7$Be ($^7$Li), equal to $[{^6}{\rm Li}((J^\pi)_T)\otimes{p}(\ell j) ]^{J^{\pi}}$ ($[{^6}{\rm Li}((J^\pi)_T)\otimes{n}(\ell j)]^{J^{\pi}}$), are built by coupling the $^6$Li wave functions with $(J^\pi)_{\rm T}$ = $1^+_1$, $1^+_2$, $3^+_1$, $0^+_1$, $2^+_1$, $2^+_2$ with the proton (neutron) wave functions in the partial waves $\ell j$: $s_{1/2}$, $p_{1/2}$, $p_{3/2}$, $d_{3/2}$, $d_{5/2}$, $f_{5/2}$, $f_{7/2}$. The cluster channels $[{^4}{\rm He}(0^+_1)\otimes{^3}{\rm He}(L_{\rm c.m.}~J_{\rm int}~J_{\rm P})]^{J^{\pi}}$ 
($[{^4}{\rm He}(0^+_1)\otimes{^3}{\rm H}(L_{\rm c.m.}~J_{\rm int}~J_{\rm P})]^{J^{\pi}}$)
are constructed by coupling $^3$He ($^3$H) wave function in partial waves: $^2S_{1/2}$, $^2P_{1/2}$, $^2P_{3/2}$, $^2D_{3/2}$, $^2D_{5/2}$, $^2F_{5/2}$, $^2F_{7/2}$, with the inert $^4$He core in $J^\pi_i = 0^+_1$ state.
Detailed information about the wave functions of low-energy $^7$Be and $^7$Li states can be seen in Tables \ref{table_channels_7Be}, \ref{table1}. 
The decomposition of low-energy wave functions of $^7$Li  and $^7$Be in the channel amplitudes and the spectroscopic factors is shown in Table \ref{table_channels_7Be}. 
Additional information is provided by the occupancies of single-particle shells for dominant configurations in the considered states (see Table \ref{table1}). 
\begin{table*}[htb]
	\caption{\label{table_channels_7Be} Major GSM-CC amplitudes of channels $^6$Li + $p$ and $^3$He + $^4$He for $^7$Be and $^6$Li + $n$ and $^3$H + $^4$He for $^7$Li. ${{\cal R}e}[b_c^2]$ denotes the real part of the channel probability ${{\cal R}e}[\braket{{\tilde w}_c|w_c}^2]$.  ${\cal S}^2$ corresponds to the GSM-CC spectroscopic factor (see Eq.~\ref{eq:spectroscopic_factor_GSM_CC}).}
	\begin{ruledtabular}
		\begin{tabular}{cccccc||cccccc}
			$^{7}$Be ; J$^{\pi}$ & $^{6}$Li ; ($J^{\pi})_{\rm T}$ & ${^3}{\rm He}$  & ${p}$ & ${{\cal R}e}[b_c^2]$ & ${\cal S}^2$ & $^{7}$Li ; J$^{\pi}$ & $^{6}$Li ; $(J^{\pi})_{\rm T}$ & ${^3}{\rm H}$  & ${n}$ & ${{\cal R}e}[b_c^2]$ & ${\cal S}^2$
			\\
			\hline 
			${ {3/2}_{1}^{-} }$  &                      & ${ ^2{P}_{3/2}}$ &                 & 0.31 & 0.04 & ${ {3/2}_{1}^{-} }$  &                  &    ${ ^2{P}_{3/2}}$ &    &              0.28  &   0.04          \\     
			                     & ${ {1}_{1}^{+} }$    &                  & ${ {p}_{3/2} }$ & 0.22 & 0.43 & & ${ {1}_{1}^{+} }$    &                  &  ${ {p}_{3/2} }$ & 0.23 & 0.44\\
			                     & ${ {1}_{1}^{+} }$    &                  & ${ {p}_{1/2} }$ & 0.10 & 0.20 & & ${ {1}_{1}^{+} }$    &                  &  ${ {p}_{1/2} }$ & 0.11 & 0.22\\	
			                     &${ {3}_{1}^{+} }$     &                  & ${ {p}_{3/2} }$ & 0.18 & 0.41 & & ${ {3}_{1}^{+} }$    &                  &  ${ {p}_{3/2} }$ & 0.18 & 0.43\\
			                     & ${ {0}_{1}^{+} }$    &                  & ${ {p}_{3/2} }$ & 0.10 & 0.22 & & ${ {0}_{1}^{+} }$    &                  & ${ {p}_{3/2} }$ & 0.10 & 0.22 \\
			                     & ${ {2}_{1}^{+} }$    &                  & ${ {p}_{1/2} }$ & 0.02 & 0.05  & & ${ {2}_{1}^{+} }$    &                  &  ${ {p}_{1/2} }$ & 0.03 & 0.08\\	
			                     &${ {2}_{2}^{+} }$     &                  & ${ {p}_{3/2} }$ & 0.03 & 0.07  &  & ${ {2}_{2}^{+} }$    &                  &  ${ {p}_{3/2} }$ & 0.03 & 0.07\\
			                     & ${ {2}_{2}^{+} }$    &                  & ${ {p}_{1/2} }$ & 0.03 & 0.08 &  & ${ {2}_{2}^{+} }$    &                  &  ${ {p}_{1/2} }$ & 0.02 & 0.07\\
			\hline          
			${ {1/2}_{1}^{-} }$  &                      & ${ ^2{P}_{1/2}}$ &                 & 0.33 &  0.04     & ${ {1/2}_{1}^{-} }$  &                 &     ${ ^2{P}_{1/2}}$ &     &             0.31    &        0.03 \\     
			                     & ${ {1}_{1}^{+} }$    &                  & ${ {p}_{3/2} }$ & 0.31 & 0.64 & & ${ {1}_{1}^{+} }$    &                  &  ${ {p}_{3/2} }$ & 0.33 & 0.67\\
                                 & ${ {1}_{1}^{+} }$    &                  & ${ {p}_{1/2} }$ & 0.06 & 0.05 & & ${ {1}_{1}^{+} }$    &                  &  ${ {p}_{1/2} }$ & 0.03 & 0.06\\
                                 & ${ {0}_{1}^{+} }$    &                  & ${ {p}_{1/2} }$ & 0.07 & 0.17 & & ${ {0}_{1}^{+} }$    &                  &  ${ {p}_{1/2} }$ & 0.08 & 0.19    \\
                                 & ${ {2}_{1}^{+} }$    &                  & ${ {p}_{3/2} }$ & 0.12 & 0.27 & & ${ {2}_{1}^{+} }$    &                  &  ${ {p}_{3/2} }$ & 0.11 & 0.23\\
			                     & ${ {2}_{2}^{+} }$    &                  & ${ {p}_{3/2} }$ & 0.06 & 0.14 & & ${ {2}_{2}^{+} }$    &                  &  ${ {p}_{3/2} }$ & 0.08 & 0.19     \\
			                     & ${ {1}_{2}^{+} }$    &                  & ${ {p}_{1/2} }$ & 0.06 & 0.12 & & ${ {1}_{2}^{+} }$    &                  &  ${ {p}_{1/2} }$ & 0.06 & 0.13\\
			\hline                    
			${ {7/2}_{1}^{-} }$  &                      & ${ ^2{F}_{7/2}}$ &                 & 0.22 & 0.03+0.02i        &  ${ {7/2}_{1}^{-} }$  &                 &     ${ ^2{F}_{7/2}}$ &     &            0.21      &      0.03+0.01i     \\    
			                     & ${ {3}_{1}^{+} }$    &                  & ${ {p}_{3/2} }$ & 0.33 & 0.70 & & ${ {3}_{1}^{+} }$    &                  &  ${ {p}_{3/2} }$ & 0.32 & 0.68\\
			                     & ${ {3}_{1}^{+} }$    &                  & ${ {p}_{1/2} }$ & 0.18 & 0.37 & & ${ {3}_{1}^{+} }$    &                  &  ${ {p}_{1/2} }$ & 0.19 & 0.40\\
			                     & ${ {2}_{1}^{+} }$    &                  & ${ {p}_{3/2} }$ & 0.07 & 0.16 & & ${ {2}_{1}^{+} }$    &                  &  ${ {p}_{3/2} }$ & 0.11 & 0.23\\
			                     & ${ {2}_{2}^{+} }$    &                  & ${ {p}_{3/2} }$ & 0.19 & 0.43 & & ${ {2}_{2}^{+} }$    &                  &  ${ {p}_{3/2} }$ & 0.17 & 0.38\\
                          \hline                    
            ${ {5/2}_{1}^{-} }$  &                      & ${ ^2{F}_{5/2}}$ &                  & 0.24 &  -0.15+0.06i    &  ${ {5/2}_{1}^{-} }$  &                &     ${ ^2{F}_{5/2}}$ &      &         0.26 &    -0.09+0.08i     \\     
			                     & ${ {1}_{1}^{+} }$    &                  &  ${ {p}_{3/2} }$ & 0.02 & 0.02-0.02i & & ${ {1}_{1}^{+} }$    &                  &  ${ {p}_{3/2} }$ & 0.01 & 0.01-0.02i   \\	
			                     & ${ {3}_{1}^{+} }$    &                  &  ${ {p}_{3/2} }$ & 0.05 & 0.11-0.07i & & ${ {3}_{1}^{+} }$    &                  &  ${ {p}_{3/2} }$ & 0.05 & 0.11-0.04i      \\
			                     & ${ {2}_{1}^{+} }$    &                  &  ${ {p}_{3/2} }$ & 0.22 & 0.45-0.11i & & ${ {2}_{1}^{+} }$    &                  &  ${ {p}_{3/2} }$ & 0.19 & 0.40-0.08i      \\	
			                     & ${ {2}_{1}^{+} }$    &                  &  ${ {p}_{1/2} }$ & 0.09 & 0.19-0.06i & & ${ {2}_{1}^{+} }$    &                  &  ${ {p}_{1/2} }$ & 0.08 & 0.14-0.03i     \\
		                         & ${ {2}_{2}^{+} }$    &                  &  ${ {p}_{3/2} }$ & 0.05 & 0.09-0.04i & & ${ {2}_{2}^{+} }$    &                  &  ${ {p}_{3/2} }$ & 0.07 & 0.13-0.04i  \\	
		                         & ${ {2}_{2}^{+} }$    &                  &  ${ {p}_{1/2} }$ & 0.10 & 0.23-0.06i & & ${ {2}_{2}^{+} }$    &                  &  ${ {p}_{1/2} }$ & 0.11 & 0.27-0.06i\\
		                         &${ {1}_{2}^{+} }$     &                  &  ${ {p}_{3/2} }$ & 0.23 & 0.47-0.12i & & ${ {1}_{2}^{+} }$    &                  &  ${ {p}_{3/2} }$ & 0.23 & 0.47-0.10i\\
			\hline
			${ {5/2}_{2}^{-} }$  &                      & ${ ^2{F}_{5/2}}$ &                  & 0.01 & -0.01+0.01i  &  ${ {5/2}_{2}^{-} }$  &                &       ${ ^2{F}_{5/2}}$ &                  & 0.01 &   -0.004+0.006i    \\ 
			                     & ${ {1}_{1}^{+} }$    &                  &  ${ {p}_{3/2} }$ & 0.46 & 0.61+0.06i & & ${ {1}_{1}^{+} }$    &                  &  ${ {p}_{3/2} }$ & 0.54 & 0.67+0.04i  \\
			                     & ${ {3}_{1}^{+} }$    &                  &  ${ {p}_{3/2} }$ & 0.18 & 0.27+0.02i & & ${ {3}_{1}^{+} }$    &                  &  ${ {p}_{3/2} }$ & 0.15 & 0.24+0.01i      \\
			                     & ${ {3}_{1}^{+} }$    &                  &  ${ {p}_{1/2} }$ & 0.27 & 0.40-0.04i & & ${ {3}_{1}^{+} }$    &                  &  ${ {p}_{1/2} }$ & 0.25 & 0.37-0.04i     \\
			                     & ${ {2}_{1}^{+} }$    &                  &  ${ {p}_{3/2} }$ & 0.06 & 0.10-0.07i & & ${ {2}_{1}^{+} }$    &                  &  ${ {p}_{3/2} }$ & 0.05 & 0.08-0.05i       \\
			                     & ${ {2}_{2}^{+} }$    &                  &  ${ {p}_{3/2} }$ & 0.01 & 0.03+0.01i & & ${ {2}_{2}^{+} }$    &                  &  ${ {p}_{3/2} }$ & 0.01 & 0.02+0.01i\\
			                     & ${ {2}_{2}^{+} }$    &                  &  ${ {p}_{1/2} }$ & 0.01 & 0.03-0.03i & & ${ {2}_{2}^{+} }$    &                  &  ${ {p}_{1/2} }$ & 0.01 & 0.02-0.02i      \\
			 \hline
			 ${ {3/2}_{2}^{-} }$ & ${ {1}_{1}^{+} }$ &                   &  ${ {p}_{3/2} }$ & 0.12 & 0.16+0.02i &      ${ {3/2}_{2}^{-} }$ & ${ {1}_{1}^{+} }$    &            &  ${ {p}_{3/2} }$ & 0.13 & 0.17+0.02i\\
			                     & ${ {1}_{1}^{+} }$    &                  &  ${ {p}_{1/2} }$ & 0.35 & 0.46+0.01i & & ${ {1}_{1}^{+} }$    &                  &  ${ {p}_{1/2} }$ & 0.35 & 0.46+0.02i\\
			                     & ${ {3}_{1}^{+} }$    &                  &  ${ {p}_{3/2} }$ & 0.04 & 0.05+0.02i & & ${ {3}_{1}^{+} }$    &                  &  ${ {p}_{3/2} }$ & 0.04 & 0.05\\	
			                     &${ {2}_{1}^{+} }$     &                  &  ${ {p}_{3/2} }$ & 0.28 & 0.42-0.05i & & ${ {2}_{1}^{+} }$    &                  &  ${ {p}_{3/2} }$ & 0.27 & 0.41-0.06i\\
			                     & ${ {2}_{1}^{+} }$    &                  &  ${ {p}_{1/2} }$ & 0.09 & 0.14-0.01i & & ${ {2}_{1}^{+} }$    &                  &  ${ {p}_{1/2} }$ & 0.09 & 0.15-0.02i\\	
			                     &${ {1}_{2}^{+} }$     &                  &  ${ {p}_{3/2} }$ & 0.12 & 0.18-0.03i & & ${ {1}_{2}^{+} }$    &                  &  ${ {p}_{3/2} }$ & 0.10 & 0.16-0.03i\\
			                     & ${ {1}_{2}^{+} }$    &                  &  ${ {p}_{1/2} }$ & 0.01 & 0.02 & & ${ {1}_{2}^{+} }$    &                  &  ${ {p}_{1/2} }$ & 0.01 & 0.01\\
                         			\hline
            ${ {1/2}_{2}^{-} }$  & ${ {1}_{1}^{+} }$    &                  &  ${ {p}_{3/2} }$ & 0.02 & 0.03+0.01i &  
            ${ {1/2}_{2}^{-} }$    &  ${ {1}_{1}^{+} }$   &             &  ${ {p}_{3/2} }$ & 0.03 & 0.03+0.01i   \\
			                     & ${ {1}_{1}^{+} }$    &                  &  ${ {p}_{1/2} }$ & 0.45 & 0.61+0.02i  & & ${ {1}_{1}^{+} }$    &                  &  ${ {p}_{1/2} }$ & 0.46 & 0.60+0.04i   \\
		                         & ${ {2}_{1}^{+} }$    &                  &  ${ {p}_{3/2} }$ & 0.11 & 0.16 & & ${ {2}_{1}^{+} }$    &                  &  ${ {p}_{3/2} }$ & 0.11 & 0.17-0.01i   \\	  
		                         & ${ {1}_{2}^{+} }$    &                  &  ${ {p}_{3/2} }$ & 0.39 & 0.60-0.07i & & ${ {1}_{2}^{+} }$    &                  &  ${ {p}_{3/2} }$ & 0.37 & 0.59-0.08i    \\
			                     & ${ {1}_{2}^{+} }$    &                  &  ${ {p}_{1/2} }$ & 0.02 & 0.03 & & ${ {1}_{2}^{+} }$    &                  &  ${ {p}_{1/2} }$ & 0.02 & 0.04-0.01i    \\
                                   \hline
            ${ {7/2}_{2}^{-} }$  & ${ {3}_{1}^{+} }$    &                  &  ${ {p}_{3/2} }$ & 0.45 & 0.61+0.05i  &    ${ {7/2}_{2}^{-} }$  & ${ {3}_{1}^{+} }$    &                  &  ${ {p}_{3/2} }$ & 0.48 & 0.66+0.06i  \\
			                     & ${ {3}_{1}^{+} }$    &                  &  ${ {p}_{1/2} }$ & 0.40 & 0.53-0.06i & & ${ {3}_{1}^{+} }$    &                  &  ${ {p}_{1/2} }$ & 0.36 & 0.50-0.04i\\
			                     & ${ {2}_{1}^{+} }$    &                  &  ${ {p}_{3/2} }$ & 0.16 & 0.23-0.07i & & ${ {2}_{1}^{+} }$    &                  &  ${ {p}_{3/2} }$ & 0.16 & 0.24-0.02i   \\
			                     & ${ {2}_{2}^{+} }$    &                  &  ${ {p}_{3/2} }$ & 0.01 & 0.001+0.003i & & ${ {2}_{2}^{+} }$    &                  &  ${ {p}_{3/2} }$ & 0.01 & 0.01+0.01i    \\
		\end{tabular}
	\end{ruledtabular}
\end{table*}
\begin{figure}[htb]
\vskip 1truecm
\begin{center}
\includegraphics[width=8cm]{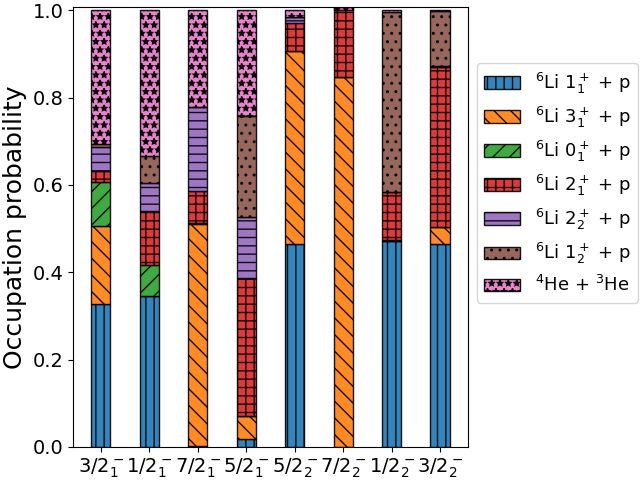}
\end{center}
\caption{(Color online) The channel decomposition for selected states of $^7$Be. In different colors, we show the summed GSM-CC probabilities of $p$ and $^3$He channels.  Detailed information about major GSM-CC channel probabilities and spectroscopic factors in individual states of $^7$Be can be found in Table \ref{table_channels_7Be}. Information about the number of protons and neutrons in different shells is given in Table \ref{table_spectra_7Be}.  }
\label{occ_7Be}
\end{figure}

Figure \ref{occ_7Be} shows the relative probability of channels [$^6$Li$((K^\pi)_{\rm T}) \otimes {p}(\ell j)]^{J^{\pi}}$ and [$^4$He$(0^+_1)\otimes {^3}{\rm He}(L_{\rm c.m.}~J_{\rm int}~J_{\rm P})]^{J^{\pi}}$ in different states of $^7$Be. 
The sum of the real parts of squared amplitudes 
${{\cal R}e}[b_c^2]\equiv{{\cal R}e}[\braket{{\tilde w}_c|w_c}^2]$ over all channels is normalized to 1, whereas the sum of imaginary parts ${{\cal I}m}[b_c^2]\equiv{{\cal I}m}[\braket{{\tilde w}_c|w_c}^2]$ is equal to 0. The basis generating one-body potential remains unchanged when the depth of a core potential varies.
One may notice that the [$^4$He$(0^+_1)\otimes {^3}{\rm He}(L_{\rm c.m.}~J_{\rm int}~J_{\rm P})]^{J^{\pi}}$ channels play a significant role only in the four lowest states: $3/2^-_1$, $1/2^-_1$, $7/2^-_1$, $5/2^-_1$ which are close to the $^4$He + $^3$He threshold. The contribution of $^3$He-cluster channel is tiny in $5/2^-_2$ resonance.
In higher lying states, the probability of this cluster channel drops below 1\%.
Notice a very different composition of wave functions in the states forming the doublet $[5/2^-_1; 5/2^-_2]$.
The mirror system $^7$Li shares a very similar channel decomposition of the states as can be deduced from Table \ref{table_channels_7Be}. 

\begin{table*}[htb]
	\caption{\label{table_spectra_7Be} Number of protons (neutrons) in resonant shells: $\ell=1$ ($p$) and non-resonant shells of the scattering continuum: $\ell=0$ ($\{s\}$), $\ell=1$ ($\{p\}$), $\ell=2$ ($\{d\}$) for largest configurations in different GSM eigenfunctions of $^{7}$Be ($^7$Li). ${{\cal R}e}[{\hat a}_c^2]$  denote the real parts of the studied squared configuration amplitudes. Only occupancies higher than 5\% are shown.} \label{table1}
	\begin{ruledtabular}
		\begin{tabular}{ccccccc||ccccccc}
			          & J$^{\pi}$   & $p$  & $\{p\}$ & $\{d\}$ & $\{s\}$ & ${{\cal R}e}[{\hat a}_c^2]$  &  & J$^{\pi}$   & $p$  & $\{p\}$ & $\{d\}$ & $\{s\}$ & ${{\cal R}e}[{\hat a}_c^2]$  \\  
			\hline 
			$^{7}$Be  & 3/2${^-_1}$ & $2p1n$  & 0         & 0         & 0         & 0.83 & $^{7}$Li  & 3/2${^-_1}$ & $1p2n$  & 0         & 0         & 0         & 0.83 \\
			          &             & $1p$    & 0         & $1p1n$      & 0         & 0.06 &  &             & $1p$    & 0         & $1p1n$      & 0         & 0.07 \\
			          & 1/2${^-_1}$ & $2p1n$  & 0         & 0         & 0         & 0.82 & & 1/2${^-_1}$ & $1p2n$  & 0         & 0         & 0         & 0.83 \\
                      & 7/2${^-_1}$ & $2p1n$  & 0         & 0         & 0         & 0.84 & & 7/2${^-_1}$ & $1p2n$  & 0         & 0         & 0         & 0.85 \\
                      & 5/2${^-_1}$ & $2p1n$  & 0         & 0         & 0         & 0.83 & & 5/2${^-_1}$ & $1p2n$  & 0         & 0         & 0         & 0.83 \\
                      & 5/2${^-_2}$ & $2p1n$  & 0         & 0         & 0         & 0.77 & & 5/2${^-_2}$ & $2p1n$  & 0         & 0         & 0         & 0.78 \\
			          &             & $1p1n$  & $1p$        & 0         & 0         & 0.09 & &             & $1p1n$  & $1n$        & 0         & 0         & 0.09 \\
			          &             & $1p$    & 0         & $1p1n$      & 0         & 0.05 & &             & $1n$    & 0         & $1p1n$      & 0         & 0.06 \\& 3/2${^-_2}$ & $2p1n$  & 0         & 0         & 0         & 0.75 & & 3/2${^-_2}$ & $1p2n$  & 0         & 0         & 0         & 0.76 \\
			          &             & $1p1n$  & $1p$        & 0         & 0         & 0.12 & &             & $1p1n$  & $1n$        & 0         & 0         & 0.11 \\
			          &             & $1p$    & 0         & $1p1n$      & 0         & 0.05 & &             & $1n$    & 0         & $1p1n$      & 0         & 0.05 \\& 1/2${^-_2}$ & $2p1n$  & 0         & 0         & 0         & 0.75 & & 1/2${^-_2}$ & $1p2n$  & 0         & 0         & 0         & 0.77 \\
			          &             & $1p1n$  & $1p$        & 0         & 0         & 0.11 &  &             & $1n$    & 0         & $1p1n$      & 0         & 0.05 \\
			          &             & $1p$    & 0         & $1p1n$      & 0         & 0.06 & &             & $1n$    & 0         & $1p1n$      & 0         & 0.05 \\
                      & 7/2${^-_2}$ & $2p1n$  & 0         & 0         & 0         & 0.77 & & 7/2${^-_2}$ & $1p2n$  & 0         & 0         & 0         & 0.78 \\
			          &             & $1p1n$  & $1p$        & 0         & 0         & 0.13 & &             & $1p1n$  & $1n$        & 0         & 0         & 0.11 \\
			          			          		\end{tabular}
	\end{ruledtabular}
	\vskip -0.6truecm
	\end{table*}

The ground state $3/2^-_1$ and the first excited state $1/2^-_1$ are dominated by the channels $[{^4}{\rm He}(0^+_1)\otimes{^3}{\rm He}(L_{\rm c.m.}~J_{\rm int}~J_{\rm P})]^{J^{\pi}}$ and $[{^6}{\rm Li}(1^+_1)\otimes{p}(p_{1/2}) ]^{J^{\pi}}$ (see Table \ref{table_channels_7Be}). 
In the $7/2^-_1$ resonance, the dominant contribution to the resonance wave function comes from the closed proton channels: $[{^6}{\rm Li}(3^+_1)\otimes{p}(p_{3/2,1/2}) ]^{7/2^-_1}$, $[{^6}{\rm Li}(2^+_2)\otimes{p}(p_{3/2}) ]^{7/2^-_1}$, and the open $^3$He channel: $[{^4}{\rm He}(0^+_1)\otimes{^3}{\rm He}(^2F_{7/2})]^{7/2^-_1}$. 

The $5/2^-_1$ resonance has still a significant component of the open channel $[{^4}{\rm He}(0^+_1)\otimes{^3}{\rm He}(^2F_{5/2})]^{5/2^-_1}$. However, the dominant contribution in the wave function of this resonance comes from the closed proton channels: $[{^6}{\rm Li}(2^+_1)\otimes{p}(p_{3/2,1/2}) ]^{5/2^-_1}$,  $[{^6}{\rm Li}(1^+_2)\otimes{p}(p_{3/2}) ]^{5/2^-_1}$. The contribution of the open proton channel $[{^6}{\rm Li}(1^+_1)\otimes{p}(p_{3/2}) ]^{5/2^-_1}$ amounts to about 2\%. 

The close lying $5/2^-_2$ resonance has a different structure than the $5/2^-_1$ resonance. The amplitude of $^3$He channel $[{^4}{\rm He}(0^+_1)\otimes{^3}{\rm He}(^2F_{5/2}) ]^{5/2^-_2}$ is only close to 1\% and the wave function is dominated by the open proton channel: $[{^6}{\rm Li}(1^+_1)\otimes{p}(p_{3/2}) ]^{5/2^-_2}$, and the closed proton channels: $[{^6}{\rm Li}(3^+_1)\otimes{p}(p_{3/2,1/2}) ]^{5/2^-_2}$. Hence, we predict that $5/2^-_1$ resonance is excited mainly in $^4$He + $^3$He reaction and decays mostly by the emission of $^3$He, whereas $5/2^-_2$ resonance is excited mainly in $^6$Li + $p$ reaction and decays predominantly by the proton emission. 

In the $7/2^-_2$ resonance, the summed probability of $[{^6}{\rm Li}(3^+_1)\otimes{p}(p_{3/2,1/2}) ]^{7/2^-_2}$ channels is 85\% and the weight of the cluster channel $[{^4}{\rm He}(0^+_1)\otimes{^3}{\rm He}(^2F_{7/2})]^{7/2^-_1}$ is totally negligible. 
In the resonances $3/2^-_2$ and $1/2^-_2$, the channels $[{^6}{\rm Li}(1^+_1)\otimes{p}(p_{3/2,1/2}) ]^{J^{\pi}}$ dominate with a summed probability of 47\%. Slightly smaller contributions come from the channels $[{^6}{\rm Li}(1^+_2)\otimes{p}(p_{3/2,1/2}) ]^{1/2^-_2}$, $[{^6}{\rm Li}(2^+_1)\otimes{p}(p_{3/2,1/2}) ]^{3/2^-_2}$ for $1/2^-_2$ and $3/2^-_2$, respectively. Other channel wave functions, including the  $^4$He + $^3$He channel, have a negligible weight in these states.

Major amplitudes of the channels $[{^4}{\rm He}(0^+_1)\otimes{^3}{\rm H}(L_{\rm c.m.}~J_{\rm int}~J_{\rm P})]^{J^{\pi}}$, $[{^6}{\rm Li}((J^\pi)_T)\otimes{n}(\ell j) ]^{J^{\pi}}$ in $^7$Li are given in Table \ref{table_channels_7Be}. 
As in $^7$Be, a significant probability of the channel wave function $[{^4}{\rm He}(0^+_1)\otimes{^3}{\rm H}(L_{\rm c.m.}~J_{\rm int}~J_{\rm P})]^{J^{\pi}}$ is seen only in the low-energy states: $J^{\pi}_i = 3/2^-_1, 1/2^-_1, 7/2^-_1, 5/2^-_1$ which are close to the $^4$He + $^3$H threshold. At higher excitation energies, the probability of the $^4$He + $^3$H channel diminishes below 1\%.

In general, mirror symmetry for low-energy wave functions in $^7$Be and $^7$Li is satisfied very well. Main probabilities are close in higher energy resonances $3/2^-_2$, $7/2^-_2$, and $1/2^-_2$ states. More significant deviations are seen for the probability of channels: $[{^6}{\rm Li}(1^+_1)\otimes{p}(p_{3/2}) ]^{5/2^-_2}$ and $[{^6}{\rm Li}(3^+_1)\otimes{p}(p_{3/2,1/2}) ]^{5/2^-_2}$ which equal 0.46 and 0.45 in $^7$Be, whereas the mirror channels in $^7$Li have the probability weights 0.54 and 0.40, respectively. 
 
Occupancies of single-particle shells in the GSM wave functions of $^7$Be and $^7$Li are very similar in all considered many-body states (see Table \ref{table1}). The probability of occupying the scattering continuum shells in $^7$Be and $^7$Li is small and does not exceed around 15\%. However, even in low-lying states: $3/2^-_1$, $1/2^-_1$, $7/2^-_1$, $5/2^-_1$, the occupation of shells in the scattering continuum, i.e.~the complement of the resonant-shell probability ($p$ only in Table \ref{table1}), amounts to about 17\% and increases to close to 25\% for higher lying states.

Table \ref{table_channels_7Be} shows also the GSM-CC spectroscopic factors for different states of $^7$Be and $^6$Li. Even though the spectroscopic factors are not observables, they nevertheless provide useful information about the configuration mixing in the many-body wave function \cite{Furnstahl2002,Duguet2015,Gomez2021,Tropiano2021}. 

In GSM-CC, the spectroscopic factor is calculated in the following
way :
\begin{eqnarray}
{\cal S}^2_{L_{\rm c.m.} J_{\rm P}} &=& \int_0^{+\infty} u_c(r)^2 \, dr \nonumber \\
&+& \left[ \sum_{N_{c.m.}} {\cal A}^2_{L_{\rm c.m.} J_{\rm P}}(N_{c.m.}) - \int_0^{+\infty} u_c(r)^2 \, dr \right]^{(\rm HO)} \label{eq:spectroscopic_factor_GSM_CC}
\end{eqnarray}
where $c$ is the channel associated to the $L_{\rm c.m.}$ and $J_{\rm P}$ quantum numbers, the superscript (HO) indicates than one projects wave functions on a basis of HO states, and ${\cal A}_{L_{\rm c.m.} J_{\rm P}}(N_{c.m.})$ is the spectroscopic amplitude :
\begin{equation} {\label{eq:spectroscopic_amplitude}}
{\cal A}_{L_{\rm c.m.} J_{\rm P}}(N_{c.m.}) = \frac{\langle \Psi_{A} || A^\dagger_{N_{c.m.} L_{\rm c.m.} J_{\rm P}} || \Psi_{A-k}\rangle}{\sqrt{2 J_A + 1}}  \ .
\end{equation}
In Eq.~(\ref{eq:spectroscopic_amplitude}), $\Psi_{A}$ and $\Psi_{A-k}$ are the wave functions of the systems with $A$ and $A-k$ nucleons, respectively. $J_A$ is the total angular momentum of the system with $A$ nucleons, and $A^\dagger_{N_{c.m.} L_{\rm c.m.} J_{\rm P}}$ is a creation operator associated with the HO projectile basis state $\ket{N_{\rm c.m.} ~ L_{\rm c.m.} ~ J_{\text{int}} ~ J_{\rm P}}$.
The integral of the squared norm of $u_c(r)$ in Eq.~(\ref{eq:spectroscopic_factor_GSM_CC}) is the GSM-CC spectroscopic factor where asymptotic properties are included exactly but antisymmetry between target and projectile is neglected. As antisymmetry is localized inside the nuclear region, it is restored by adding the GSM spectroscopic factor projected in a HO basis, where antisymmetry is exactly taken into account via the use of Slater determinants, and by removing the squared norm of the HO projected $u_c(r)$ channel wave function.

One may see in Table \ref{table_channels_7Be} a qualitative agreement between the real parts of the one-proton channel probability ${\cal R}e[b_c^2]$ and the corresponding real parts of the spectroscopic factor, i.e. large probabilities for the channels: $\{ [{^6}{\rm Li}((K^\pi))\otimes{p}(\ell j) ]^{J^{\pi}} \}^2$, correspond to large real parts of the spectroscopic factors: $^7$Be$(J^{\pi}) \rightarrow {p}(\ell j) \oplus {^6}{\rm Li}((K^\pi))$. Similarly for $^7$Li, there is a close qualitative relation between magnitudes of  probabilities ${\cal R}e[b_c^2]$ for channels: $\{ [{^6}{\rm Li}((K^\pi))\otimes{n}(\ell j) ]^{J^{\pi}} \}^2$ and values of the spectroscopic factors ${\cal R}e[{\cal S}^2]$: $^7$Li$(J^{\pi}) \rightarrow {n}(\ell j) \oplus {^6}{\rm Li}((K^\pi))$. However, the one-nucleon spectroscopic factors in mirror configurations of $^7$Be and $^7$Li may show in some cases deviations up to $\approx 20\%$.

One may also notice that the real parts of cluster spectroscopic factors in $^7$Be and $^7$Li are significantly smaller than the dominant one-nucleon spectroscopic factors, even though the largest one-nucleon channel probabilities are of the same order as the cluster channel probabilities. Moreover, in the doublet of resonances $[5/2^-_1;5/2^-_2]$, the cluster spectroscopic factors are small and negative. 
It is to be noted that in the complex-energy framework such as GSM or GSM-CC, the spectroscopic factor ${\cal R}e[{\cal S}^2]$ or the squared wave-function amplitude $\mathcal{R}e[b_c^2]$  can be negative in resonance states \cite{michel_review_2009}. The statistical uncertainty of ${\cal R}e[{\cal S}^2]$ 
is, at the leading order, associated with its imaginary part ${\cal I}m[{\cal S}^2]$ 
\cite{Berggren1996,michel_book_2021}. A statistical uncertainty on ${\cal I}m[{\cal S}^2]$ 
arises because of the different life times that the resonance state can have in several experiments. Hence, as explained in 
Refs.\,\cite{Berggren1996,michel_book_2021,Myo2023}, ${\cal R}e[{\cal S}^2]$ 
is the average value of the corresponding spectroscopic factor 
obtained in different measurements, while ${\cal I}m[{\cal S}^2]$ 
can be related to the dispersion rate over time in the measurement, and hence represents its statistical uncertainty. In the case of the doublet of resonances: $[5/2^-_1;5/2^-_2]$, an imaginary part of the spectroscopic factor is of the same order of magnitude as the real part $\mathcal{R}e[{\cal S}^2]$. Thus, there is a large statistical uncertainty on cluster spectroscopic factors in these resonances.

\subsection{Reaction cross sections } \label{crosssec}
In this section, we shall discuss the reaction cross-sections $^4$He($^3$He, $^3$He) , $^4$He($^3$H, $^3$H), and $^6$Li($p,p$). The GSM-CC cross sections are calculated by coupling the real-energy incoming partial waves to the states of $^4$He or $^{6}$Li given by the Hermitian Hamiltonian. GSM-CC calculations are performed using COSM coordinates \cite{Suzuki1988} but 
the reaction cross sections will be expressed in the c.m.~reference frame. The initial energy is $E^{\rm (COSM)}=E_{\rm proj}^{\rm (COSM)} + E_{\rm T}^{\rm (COSM)}$, where $E^{\rm (COSM)}$, $E_{\rm proj}^{\rm (COSM)}$, and $E_{\rm T}^{\rm (COSM)}$ are the total energy, the projectile energy, and the GSM target binding energy, respectively. One has $E_{\rm proj}^{\rm (lab)} \simeq 1.25 ~ E_{\rm proj}^{\rm (COSM)}$ for $^4$He($^3$H,$^3$H)$^4$He and $^4$He($^3$He,$^3$He)$^4$He reactions and $E_{\rm proj}^{\rm (c.m.)} \simeq 1.07 ~ E_{\rm proj}^{\rm (COSM)}$ for the $^6$Li($p,p$)$^6$Li reaction \cite{michel_book_2021}.

\textit{Reaction}:~$^4$He($^3$H,$^3$H)$^4$He~--- 
Figure \ref{3H4Hecsec} shows the $^4$He($^3$H,$^3$H)$^4$He differential cross section calculated in GSM-CC (solid line). Triton bombarding energy is given in the laboratory frame. The calculation is performed using the same Hamiltonian and the same model space as used in the calculation of the spectra of $^{6,7}$Li. $^3$H energies in Fig.~\ref{3H4Hecsec} are in the c.m.~reference frame. The estimated error on experimental results varies with the $^3$H bombarding energy and amounts to about 10\% \cite{spiger1967scattering}. 
\begin{figure}[htb]
\begin{center}
\includegraphics[width=8cm]{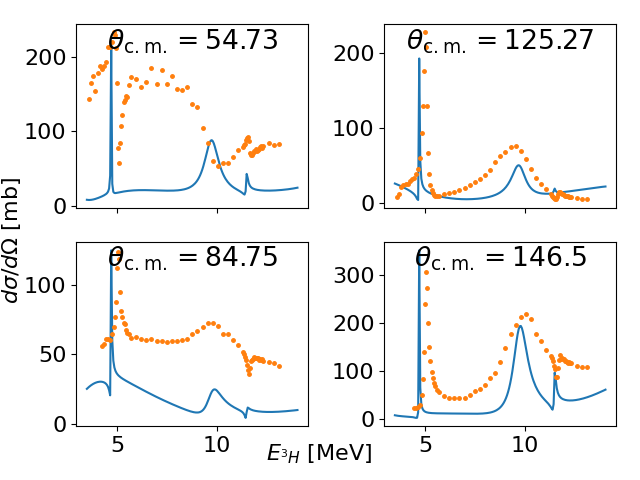}
\end{center}
\caption{(Color online) The GSM-CC elastic differential cross sections of the reaction $^{4}$He($^{3}$H,$^{3}$H)$^{4}$He, calculated at four different c.m.~angles, are compared with the experimental data (in dots) \cite{spiger1967scattering}. $^3$H bombarding energy is given in the laboratory frame.}
\label{3H4Hecsec}
\end{figure}
Peaks in the calculated cross section correspond to $7/2^-_1$, $5/2^-_1$, and $5/2^-_2$ resonances (see Fig.~\ref{spec_7Li7Be}). The agreement between theory and experiment is satisfactory at backward angles whereas at forward angles the GSM-CC underestimates experimental cross-sections. However, the reason for this underestimation of cross section at forward angles has not been identified. \\

\textit{Reaction}:~$^4$He($^3$He,$^3$He)$^4$He~--- 
Figure \ref{3He4Hecsec} shows the $^4$He($^3$He,$^3$He)$^4$He elastic differential cross section calculated in GSM-CC (solid line) at different c.m.~angles.
\begin{figure}[htb]
\begin{center}
\includegraphics[width=8cm]{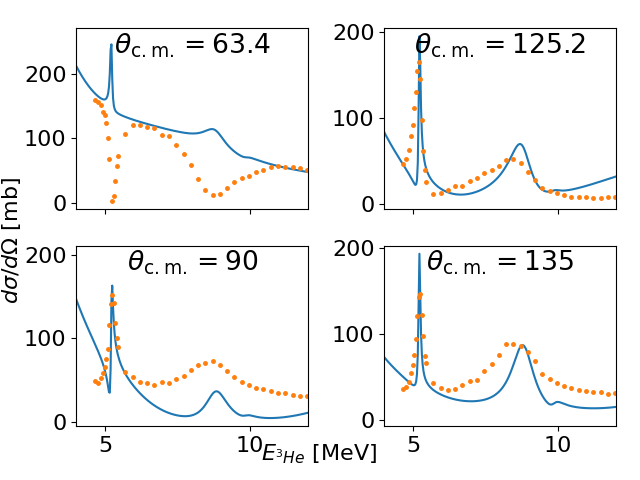}
\end{center}
\caption{(Color online) The same as in Fig.~\ref{3H4Hecsec} but for the reaction $^{4}$He($^{3}$He,$^{3}$He)$^{4}$He. Experimental data (in dots) are from Ref.\cite{spiger1967scattering}.}
\label{3He4Hecsec}
\end{figure}
The GSM-CC cross section provides a good description of experimental cross-sections at large angles. At $\Theta_{\rm c.m.}\leq 90$ $\deg$, the calculated cross sections underestimate the experimental ones.  \\

\textit{Reaction}:~$^6$Li($p,p$)$^6$Li~--- 
Figure \ref{6Lippcseccsec} shows the proton elastic differential cross section $^6$Li($p,p$)$^6$Li calculated in GSM-CC (solid line) at different c.m.~angles.
\begin{figure}[htb]
\begin{center}
\includegraphics[width=8cm]{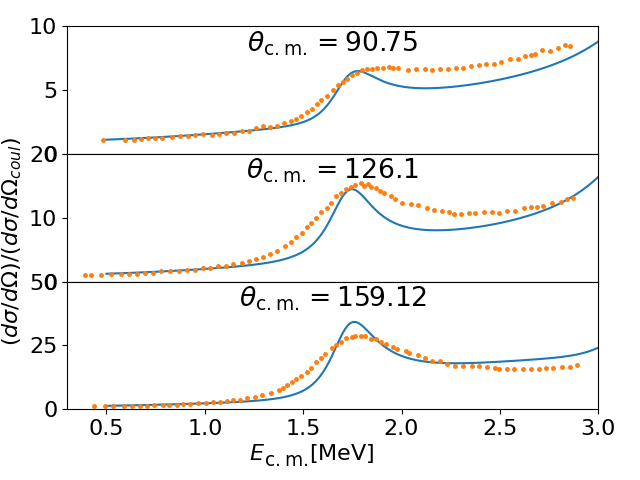}
\end{center}
\caption{The GSM-CC elastic differential cross sections of the reaction {$^6${\rm Li}($p,p$)$^6${\rm Li}} at three different c.m.~angles are plotted as a function of proton energy and compared with experimental data (in dots) \cite{mccray1963elastic}. Proton bombarding energy is given in the c.m.~frame. Both GSM-CC and experimental cross sections are divided by the Rutherford cross section.}
\label{6Lippcseccsec}
\end{figure}
Experimental results are well reproduced by the GSM-CC calculation.

\subsection{Near-threshold behavior of the channel amplitudes and spectroscopic factors}
\label{nearthrsec}
As the incident energy increases and new reaction channels open up, the reaction threshold becomes a point of bifurcation for particle flux. Due to the unitarity of the scattering matrix and the resulting flux conservation, an opening reaction channel can generate changes in other open channels. 

The appearance and properties of near-threshold cusps has been explained by Wigner who formulated the threshold law for the elastic and total cross sections. Later this phenomenon has found explanation in terms of the R-matrix theory \cite{Breit1957,Baz1957,Newton1958,Fonda1961,Meyerhof1963,Baz1969,Lane1970}. Wigner cusps are the non-analyticities appearing in cross sections at energies when a new particle emission channel opens. They translate into very sharp changes of cross sections, discontinuities of the derivative of cross sections or of the cross sections themselves, depending on the angular momentum content of the wave functions involved in cross sections.

Another facet of this phenomenon has been demonstrated in the GSM study of near-threshold behavior of the spectroscopic factors \cite{Michel2007,Michel2007a} which exhibit qualitatively different features according to the bound or unbound nature of involved eigenstates. Wigner cusps appear in spectroscopic factors when a many-body state crosses particle emission threshold. They are particularly well visible for neutron $\ell = 0,1$ waves, while they occur only in spectroscopic factor derivatives in neutron waves with $\ell \geq 2$. No Wigner cusp  appears in spectroscopic factors for charged particles. 

Spectroscopic factors measure occupancy of the single-particle shells and, hence the role of nucleon-nucleon correlations. Observation of near-threshold irregularities in spectroscopic factors raise the question how the proximity of particle-emission threshold and, in particular, coupling to non-resonant continuum, changes the structure of nuclear states. Coupling to the non-resonant scattering continuum in GSM is essential to correctly describe the energy-dependence of spectroscopic factors close to particle-emission threshold. Neglecting this coupling removes a singular behavior in the energy-dependence of spectroscopic factors, i.e.~coupling to the non-resonant scattering continuum is crucial to preserve the unitarity at particle-emission threshold. Therefore, salient features of both  cross-sections and  spectroscopic factors for states in the vicinity of decay thresholds are  direct consequences of unitarity in near-threshold wave functions. Another effect of unitarity is the breaking of isospin symmetry in mirror nuclei due to the very different asymptotic behavior of proton and neutron wave functions at particle-emission threshold \cite{PRC_isospin_mixing}. Hence, continuum coupling may indeed lead to isospin breaking even in the absence of Coulomb interaction. 

The genuine near-threshold properties of quantum states depend on the nature of configuration mixing between resonant states and the continuum of scattering states. It was conjectured that an interplay between Hermitian and anti-Hermitian continuum couplings leads to the concentration of the collective strength in a single state, the so-called aligned state of an open quantum system, which shares many properties of a nearby decay threshold \cite{Okolowicz2012,Okolowicz2013,Okolowicz2018,Ploszajczak2020,Okolowicz2020}. 
Favorable conditions for the formation of the aligned eigenstate have been investigated in 
the shell model embedded in the continuum (SMEC) \cite{Bennaceur99,Bennaceur00,Okolowicz2003,ROTUREAU2006}, by calculating the correlation energy due to the continuum coupling. The GSM in Slater determinant representation is not a proper tool for a study of the alignment effect in near-threshold states because the correct asymptotic of many-body eigenstates and decay channels cannot be imposed in GSM. On the contrary, GSM-CC is well suited for that matter, as the latter nuclear state properties are well defined therein. Below, we shall present selected examples of the alignment phenomenon in the eigenstates of $^7$Li, $^7$Be, and $^8$Be, which can be seen in the near-threshold dependence of the reaction channel probabilities.

\subsection{Energy dependence of channel amplitudes and spectroscopic factors in $^7$Li and $^7$Be} \label{nearthrA=7}

\subsubsection{Channel probabilities and resonance widths}
As discussed in Sec.~\ref{spectra7} and seen in Fig.~\ref{spec_7Li7Be}, the lowest decay threshold in $^7$Li is the $^3$H cluster thresholds. 
Neutron decay thresholds $[{^6}{\rm Li}(K^\pi)\otimes{n} (\ell j)]^{J^{\pi}}$ open at higher excitation energies. 
\begin{figure}[htb]
\begin{center}
\includegraphics[width=8.5cm]{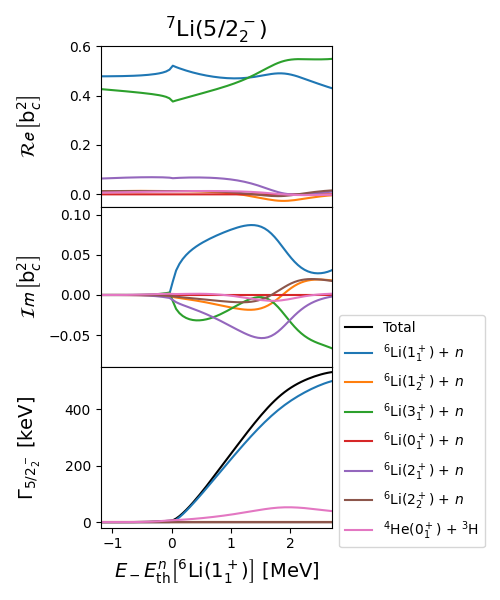}
\end{center}
\caption{(Color online) From top to bottom: the real and imaginary parts of the squared amplitudes $b_c^2\equiv\braket{{\tilde w}_c|w_c}^2$, and the partial widths of each reaction channel in the $5/2^-_2$ state of  $^7$Li. Contributions of all partial waves $\ell j$ in reaction channels $[{^6}{\rm Li}(1^+_1)\otimes{n}(\ell j)]^{5/2^-}$ are summed and everything is plotted as a function of the distance of the state to the neutron threshold. The meaning of curves for different channels is explained in the attached insert. Each curve represents the sum of all reaction channels built on the same many-body state of $^6$Li.}
\label{7Li5|2-_1}
\end{figure}

The dependence of channel probabilities and partial widths $\Gamma_c$ in the $5/2^-_2$ state of $^7$Li on the energy difference with respect to the lowest one-neutron decay threshold $[{^6}{\rm Li}(1^+_1)\otimes{n}(\ell j)]^{J^{\pi}}$ is shown in Fig. \ref{7Li5|2-_1}. Both real and imaginary parts of the channel probabilities are shown. The partial widths are calculated using the current formula \cite{barmore2000theoretical}. The energy difference between the $5/2^-_2$ state and neutron-threshold energy is varied by changing the depth $V_0$ of the $^4$He core potential. We also calculate $^6$Li with the same interaction. This means that the energy of $^6$Li will move alongside the energy of $^7$Li and $^7$Be, however, the binding energy of the clusters $^3$H, $^3$He in $^7$Li, $^7$Be, respectively, remains unchanged.

For $E-E_{\rm th}^n[^6{\rm Li}(1_1^+)]<0$, all neutron reaction channels are closed, i.e. the state $5/2_2^-$ is bound with respect to the emission of neutron. At $E-E_{\rm th}^n[^6{\rm Li}(1_1^+)]\simeq -0.8$ MeV, the triton channel $[{^4}{\rm He}(0^+_1)\otimes{^3{\rm H}}(^2F_{5/2})]^{5/2^{-}}$ opens. This opening has no visible consequences for the wave function 
\begin{figure}[htb]
\begin{center}
\includegraphics[width=8.5cm]{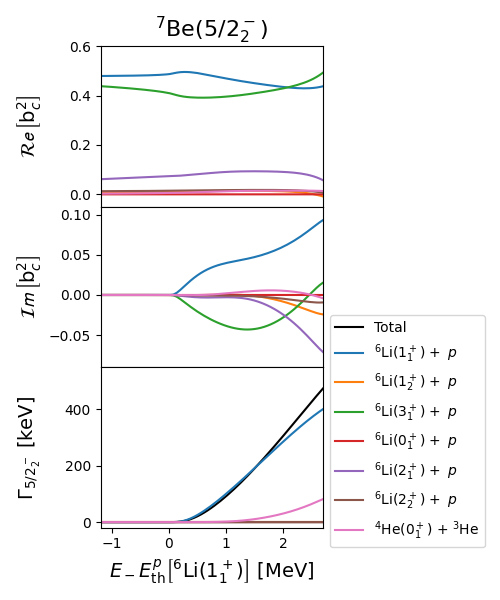}
\end{center}
\caption{(Color online) Same as Fig. \ref{7Li5|2-_1} 
but for the mirror system $^7$Be(5/2$^-_2$). 
Everything is plotted as a function of the 
distance of the state to the proton 
threshold.}
\label{7Be5|2-_1}
\end{figure}
$5/2^-_2$. For $E-E_{\rm th}^n[^6{\rm Li}(1_1^+)] > 0$, both neutron and triton can be emitted.  

The contribution of the reaction channel $[{^6}{\rm Li}(1^+_1)\otimes{n}(\ell j)]^{5/2^-}$ dominates below neutron emission threshold. The second largest probability corresponds to the channel $[{^6}{\rm Li}(3^+_1)\otimes{n}(\ell j)]^{5/2^-}$. In the vicinity of the neutron threshold, the probability of a channel $[{^6}{\rm Li}(1^+_1)\otimes{n}(\ell j)]^{5/2^-}$ promptly increases, what is a manifestation of an alignment of the  GSM-CC state $5/2_2^+$ with the neutron decay channel $[{^6}{\rm Li}(1^+_1)\otimes{n}(\ell j)]^{J^{\pi}}$.  One may notice a Wigner cusp in the energy dependence of a   probability of the channel $[{^6}{\rm Li}(1^+_1)\otimes{n}(\ell j)]^{5/2^-}$. The increased probability of this channel is associated with the opposite effect in other channels, mainly in the channel $[{^6}{\rm Li}(3^+_1)\otimes{n}(\ell j)]^{5/2^-}$.
At higher energies below opening of the next neutron channel at $E-E_{\rm th}^n[^6{\rm Li}(1_1^+)] \simeq 2.93$ MeV, weight of the closed channel 
$[{^6}{\rm Li}(3^+_1)\otimes{n}(\ell j)]^{5/2^-}$ increases gradually and this channel becomes dominant $\approx$ 1 MeV above the neutron threshold $[{^6}{\rm Li}(1^+_1)\otimes{n}(\ell j)]^{J^{\pi}}$. 

The imaginary parts of squared amplitudes ${\cal I}m[b_c^2]$
are zero below the neutron emission threshold and show a complicate behavior above it. Above the decay threshold 
$[{^6}{\rm Li}(1^+_1)\otimes{n}(\ell j)]^{J^{\pi}}$ and below the next threshold $[{^6}{\rm Li}(3^+_1)\otimes{n}(\ell j)]^{J^{\pi}}$, the major contribution comes from the open reaction channel $[{^6}{\rm Li}(1^+_1)\otimes{n}(\ell j)]^{5/2^-}$. The magnitude of this contribution decreases when we approach the threshold of the channel $[{^6}{\rm Li}(3^+_1)\otimes{n}(\ell j)]^{J^{\pi}}$ whereas the magnitude of the (negative) contribution of the channel  $[{^6}{\rm Li}(3^+_1)\otimes{n}(\ell j)]^{5/2^-}$ strongly increases and changes sign at the threshold. The interplay between these two reaction channels continues to dominate the evolution pattern of imaginary parts above this threshold. 

\begin{figure}[htb]
\begin{center}
\includegraphics[width=8.5cm]{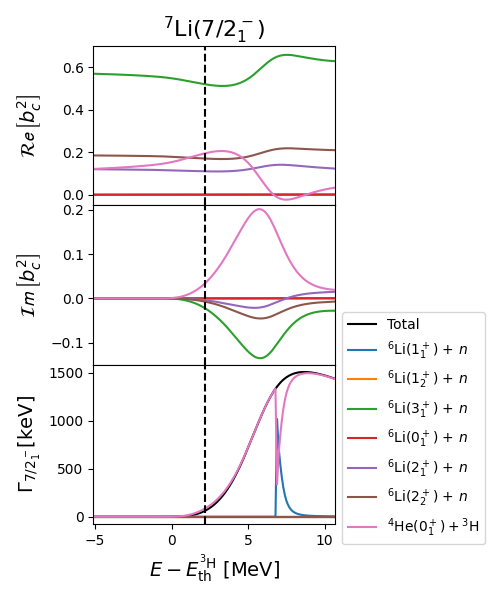}
\end{center}
\caption{(Color online) The real and imaginary parts of the channel probabilities $b_c^2$ and the partial widths of each reaction channel in the $7/2^-_1$ state of $^7$Li are shown as a function of the distance with respect to the lowest decay threshold $[{^4}{\rm He}(0^+_1)\otimes{^3{\rm H}}(L_{\rm c.m.}~J_{\rm int}~J_{\rm P})]^{J^{\pi}}$. The dashed vertical line gives the GSM-CC energy of the $7/2^-_1$ resonance. For more details, see the caption of Fig.~\ref{7Li5|2-_1} and discussion in the text.
 }
\label{7Li7|2^-_1}
\end{figure}

Figure \ref{7Be5|2-_1} presents the dependence of channel probabilities and partial widths in the $5/2^-_2$ state of $^7$Be on the energy difference with respect to the lowest one-proton decay threshold $[{^6}{\rm Li}(1^+_1)\otimes{p}(\ell j)]^{J^{\pi}}$. In this mirror state of a $5/2^-_2$ state in $^7$Li, 
opening of the $^3$He channel $[{^4}{\rm He}(0^+_1)\otimes{^3{\rm He}}(^2F_{5/2})]^{5/2^{-}}$ 
has no visible consequences for a structure of the $5/2^-_2$ wave function. $^3$He can be emitted, though a $^3$He decay width becomes significant only for $E-E_{\rm th}^p[^6{\rm Li}(1_1^+)] > 1.5$ MeV.

Below the proton emission threshold, the 
contribution of a reaction channel $[{^6}{\rm Li}(1^+_1)\otimes{p}(\ell j)]^{5/2^-}$ dominates. The second largest probability corresponds to the channel $[{^6}{\rm Li}(3^+_1)\otimes{p}(\ell j)]^{5/2^-}$. In the vicinity of the proton threshold, the probability of a channel $[{^6}{\rm Li}(1^+_1)\otimes{p}(\ell j)]^{5/2^-}$ slightly increases, what is a manifestation of the alignment with the proton decay channel. 
The Coulomb interaction is smoothing this effect. The increased probability of the channel $[{^6}{\rm Li}(1^+_1)\otimes{p}(\ell j)]^{5/2^-}$ is associated with 
\begin{figure}[htb]
\begin{center}
\includegraphics[width=8.5cm]{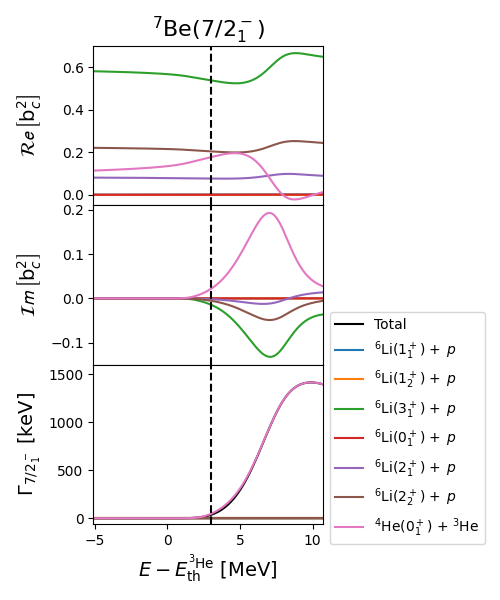}
\end{center}
\caption{(Color online) Same as Fig. \ref{7Li7|2^-_1} 
but for the mirror system $^7$Be(7/2$^-_1$). 
Everything is plotted as a function of the 
distance of the state to the $^3$He threshold.}
\label{7Be7|2-_1}
\end{figure}
the opposite effect in the channel $[{^6}{\rm Li}(3^+_1)\otimes{p}(\ell j)]^{5/2^-}$.
At higher energies, below an opening of the next proton channel, 
the probability of the closed channel 
$[{^6}{\rm Li}(3^+_1)\otimes{p}(\ell j)]^{5/2^-}$ increases gradually and dominates at $E-E_{\rm th}^p[^6{\rm Li}(1_1^+)] \simeq 2$ MeV. 

The imaginary parts of squared amplitudes 
${\cal I}m[b_c^2]$ are different from zero above the proton emission threshold. The major contribution comes from the open reaction channel $[{^6}{\rm Li}(1^+_1)\otimes{p}(\ell j)]^{5/2^-}$. The magnitude of this contribution does not decrease even in the vicinity of the next threshold $[{^6}{\rm Li}(3^+_1)\otimes{n}(\ell j)]^{J^{\pi}}$. 

Figure \ref{7Li7|2^-_1} presents the dependence of channel probabilities and partial widths $\Gamma_c$ on the energy difference with respect to the lowest decay threshold $[{^4}{\rm He}(0^+_1)\otimes{^3{\rm H}}(L_{\rm c.m.}~J_{\rm int}~J_{\rm P})]^{J^{\pi}}$ in the $7/2^-_1$ state of $^7$Li. 
In the whole interval of energies, neutron reaction channel $[{^6}{\rm Li}(3^+_1)\otimes{n}(\ell j)]^{7/2^-}$ provides the major contribution. The probability of the $^3$H channel $[{^4}{\rm He}(0^+_1)\otimes{^3{\rm H}}(^2F_{7/2})]^{7/2^-}$ grows when approaching the emission threshold and has a maximum close to 3.3 MeV above the threshold energy $E_{\rm th}^{{^{3}H}}$. The maximum of the imaginary part of $^3$H probability is shifted to slightly higher energy with respect to the maximum of the real part.

The alignment of the $7/2^-_1$ state with the decay channel $[{^4}{\rm He}(0^+_1)\otimes{^3{\rm H}}(^2F_{7/2})]^{7/2^{-}}$ is shifted at a higher energy due to a combined effect of the Coulomb interaction and the angular momentum involved in this reaction channel. This alignment is accompanied by the anti-alignment with respect to the $[{^6}{\rm Li}(3^+_1)\otimes{n}(\ell j)]^{7/2^-}$ decay channel.
Above the neutron decay threshold at $E-E_{\rm th}^{{^3{\rm H}}}[^6{\rm Li}(1_1^+)] \simeq 6.8$ MeV in the channel $[{^6}{\rm Li}(1^+_1)\otimes{n}(\ell j)]^{7/2^-}$, which is barely visible in Fig.~\ref{7Li7|2^-_1}, the squared amplitude of the $^3$H channel strongly diminishes and the closed channel $[{^6}{\rm Li}(3^+_1)\otimes{n}(\ell j)]^{7/2^-}$ becomes dominant again in the wave function of $7/2^-_1$ state. This radical change of the $7/2^-_1$ eigenstate in the vicinity of the neutron threshold is seen in the energy dependence of decay probability, which both below and above neutron emission threshold corresponds to $^3$H emission. However, one can see an interplay between neutron and $^3$H channels in a narrow range of energies at around the decay threshold. 

The mirror system to $7/2^-_1$ of $^7$Li is shown in Fig.  \ref{7Be7|2-_1} which presents the dependence of channel probabilities and partial widths in the $7/2^-_1$ state of $^7$Be on the energy difference with respect to the lowest threshold $[{^4}{\rm He}(0^+_1)\otimes{^3{\rm He}}(L_{\rm c.m.}~J_{\rm int}~J_{\rm P})]^{J^{\pi}}$. In spite of different Coulomb interaction in mirror $7/2^-_1$ states, one may notice large similarity of the calculated energy dependencies shown in Figs. \ref{7Li7|2^-_1} and \ref{7Be7|2-_1}. 

\subsubsection{Near-threshold energy dependence of the spectroscopic factors} 
In this section we discuss the near-threshold effects in the spectroscopic factors calculated using the GSM- CC wave functions. 
\begin{figure}[htb]
\begin{center}
\includegraphics[width=8.5cm]{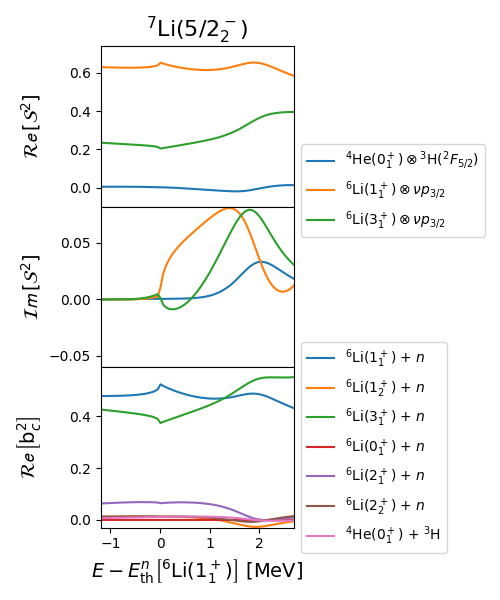}
\end{center}
\caption{(Color online) The spectroscopic factors and the channel weights in $5/2^-_2$ state of $^7$Li are shown as a function of the distance with respect to the neutron emission threshold $[{^6}{\rm Li}(1^+_1)\otimes{n}(\ell j)]^{J^{\pi}}$. From top to bottom: (i) real part of the spectroscopic factors ${\cal R}e[{\cal S}^2]$,   (ii) imaginary part of the spectroscopic factors ${\cal I}m[{\cal S}^2]$,  and (iii) real part of the channel weights ${\cal R}e[b_c^2]$. 
 }
\label{7Li5|2^-_2_SF}
\end{figure}
\begin{figure}[htb]
\begin{center}
\includegraphics[width=8.5cm]{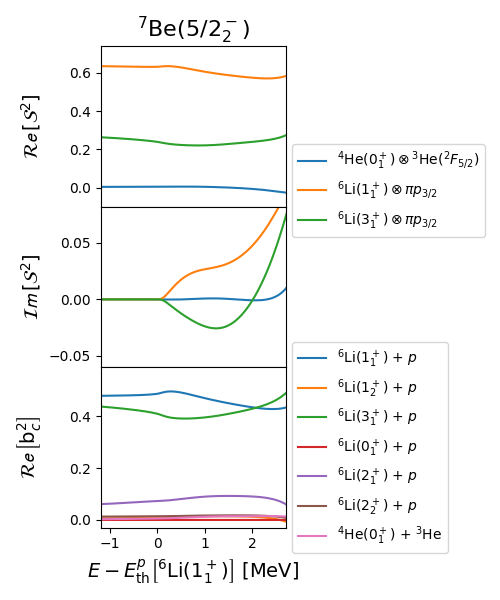}
\end{center}
\caption{(Color online) Same as 
Fig. \ref{7Li5|2^-_2_SF} but for the mirror system $^7$Be(5/2$^-_2$). the spectroscopic factors and the channel weights ${\cal R}e[b_c^2]$ are shown as a function of the distance with respect to the one-proton decay threshold $^6{\rm Li}(1^+_1)\otimes{p}(\ell j)]^{5/2^-}$.
 }
\label{7Be5|2^-_2_SF}
\end{figure}
Figure \ref{7Li5|2^-_2_SF} shows the real and imaginary parts of the spectroscopic factors and the real part of the channel probabilities in $5/2^-_2$ resonance of $^7$Li. Only largest neutron spectroscopic factors are shown. All quantities are plotted as a function of the energy difference with respect to the lowest neutron emission threshold 
$[{^6}{\rm Li}(1^+_1)\otimes{n}(\ell j)]^{J^{\pi}}$. As expected, one may notice a Wigner cusp in the spectroscopic factor $\langle ^7{\rm Li}(5/2^-_2)|[^6{\rm Li}(1^+_1)\otimes{n}(p_{3/2})]^{5/2^-}\rangle^2$ at the threshold which mimics the cusp in the probability of a channel $[{^6}{\rm Li}(1^+_1)\otimes{n}(\ell j)]^{5/2^-}$. This local increase of the spectroscopic factor $\langle ^7{\rm Li}(5/2^-_2)|[^6{\rm Li}(1^+_1)\otimes{n}(p_{3/2})]^{5/2^-}\rangle^2$ is accompanied by the slight decrease of the spectroscopic factor $\langle ^7{\rm Li}(5/2^-_2)|[^6{\rm Li}(3^+_1)\otimes{n}(p_{1/2})]^{5/2^-}\rangle^2$. The same feature is seen in the channel probabilities ${\cal R}e[b_c^2]$.
In the whole range of energies represented in Fig. \ref{7Li5|2^-_2_SF}, the spectroscopic factor $\langle ^7{\rm Li}(5/2^-_2)|[^6{\rm Li}(1^+_1)\otimes{n}(p_{3/2})]^{5/2^-}\rangle^2$ dominates but the difference with the value of the second largest spectroscopic factor $\langle ^7{\rm Li}(5/2^-_2)|[^6{\rm Li}(3^+_1)\otimes{n}(p_{1/2})]^{5/2^-}\rangle^2$ diminishes when the energy approaches the threshold of a channel $[{^6}{\rm Li}(3^+_1)\otimes{n}(\ell j)]^{5/2^-}$.

The imaginary part ${\cal I}m[{\cal S}^2]$ of the spectroscopic factor $\langle ^7{\rm Li}(5/2^-_2)|[^6{\rm Li}(1^+_1)\otimes{n}(p_{3/2})]^{5/2^-}\rangle^2$ starts to grow above the threshold and dominates until $E-E_{\rm th}^n[^6{\rm Li}(1_1^+)]\approx 1.6$ MeV when it is surpassed by the spectroscopic factor in the channel $[{^6}{\rm Li}(3^+_1)\otimes{n}(\ell j)]^{5/2^-}$.

Results for a mirror $5/2^-_2$ state in $^7$Be are shown in Fig. \ref{7Be5|2^-_2_SF}. The spectroscopic factor $\langle ^7{\rm Be}(5/2^-_2)|[^6{\rm Li}(1^+_1)\otimes{p}(p_{3/2})]^{5/2^-}\rangle^2$ 
is changing smoothly around the proton threshold 
$[{^6}{\rm Li}(1^+_1)\otimes{p}(\ell j)]^{J^{\pi}}$. The imaginary part of $\langle ^7{\rm Be}(5/2^-_2)|[^6{\rm Li}(1^+_1)\otimes{p}(p_{3/2})]^{5/2^-}\rangle^2$ grows gradually and at energies $E-E_{\rm th}^n[^6{\rm Be}(1_1^+)]> 2.75$ MeV is approached by the spectroscopic factor $[{^6}{\rm Li}(3^+_1)\otimes{n}(\ell j)]^{5/2^-}$.

\begin{figure}[htb]
\begin{center}
\includegraphics[width=8.5cm]{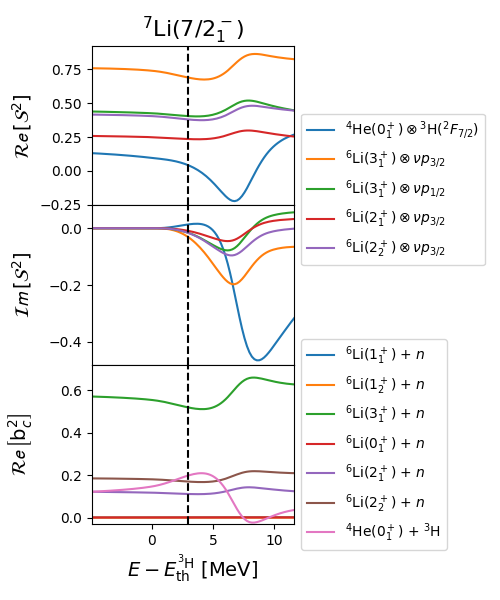}
\end{center}
\caption{(Color online) The spectroscopic factors and the channel weights in $7/2^-_1$ state of $^7$Li are shown as a function of the distance with respect to the triton emission threshold $[{^4}{\rm He}(0^+_1)\otimes{^3{\rm H}}(L_{\rm c.m.}~J_{\rm int}~J_{\rm P})]^{J^{\pi}}$. From top to bottom: (i) real part of the spectroscopic factors ${\cal R}e[{\cal S}^2]$,   (ii) imaginary part of the spectroscopic factors ${\cal I}m[{\cal S}^2]$,  and (iii) real part of the channel weights ${\cal R}e[b_c^2]$. The dashed vertical line gives the GSM-CC energy of the $7/2^-_1$ resonance.
 }
\label{7Li7|2^-_1_SF}
\end{figure}
Figure \ref{7Li7|2^-_1_SF} shows the real and imaginary parts of the spectroscopic factors and the real part of the channel probabilities in $7/2^-_1$ resonance of $^7$Li. Only largest neutron spectroscopic factors are shown. All quantities are plotted as a function of the energy difference with respect to the triton emission threshold 
$[{^4}{\rm He}(1^+_1)\otimes {^3{\rm H}}(L_{\rm c.m.} J_{\rm int} J_{\rm P})]^{J^{\pi}}$. One may notice that the energy dependence of the triton $\langle ^7{\rm Li}(7/2^-_1)|[^4{\rm He}(0^+_1)\otimes {^3{\rm H}}(^1F_{7/2})]^{7/2^{-}}\rangle^2$ and one-neutron spectroscopic factors resemble the dependencies seen in the channel probabilities ${\cal R}e[b_c^2]$. The minimum of the real part of the triton spectroscopic factor is negative with a large uncertainty associated with the imaginary part. Interestingly, the triton spectroscopic factor increases and becomes positive when approaching the one-neutron emission threshold $[{^6}{\rm Li}(3^+_1)\otimes{p}(\ell j)]^{J^{\pi}}$. In the whole range of energies represented in Fig. \ref{7Li7|2^-_1_SF}, the dominant spectroscopic factors are: $\langle ^7{\rm Li}(7/2^-_1)|[^6{\rm Li}(3^+_1)\otimes{n}(p_{3/2,1/2})]^{7/2^-}\rangle^2$, 
$\langle ^7{\rm Li}(7/2^-_1)|[^6{\rm Li}(2^+_1)\otimes{n}(p_{3/2})]^{7/2^-}\rangle^2$, 
$\langle ^7{\rm Li}(7/2^-_1)|[^6{\rm Li}(2^+_2)\otimes{n}(p_{3/2})]^{7/2^-}\rangle^2$.

\section{Conclusions}
\label{conclusions}
{
We have applied in this work the multi-mass-partition GSM-CC approach for the description of $^7$Li and $^7$Be. The lowest threshold in these nuclei corresponds to the emission of clusters of nucleons and therefore the GSM-CC description of resonance wave functions requires the inclusion of the reaction channels involving both clusters and nucleons. With the two mass partitions $^6$Li + $n$, $^4$He + $^3$H for $^7$Li and $^6$Li + $p$, and $^4$He + $^3$He for $^7$Be, we obtain a good description of resonance energies and widths in the coupled channel framework of the GSM-CC. In the same framework, we have calculated elastic scattering $^4$He + $^3$H, $^4$He + $^3$He, and $^6$Li + $p$. The GSM-CC provides a reasonable description of elastic differential cross-sections in these reactions, particularly at backward angles.

Generally, the calculated resonance widths are somewhat smaller than found experimentally. Small real-energy correction factors of the channel-channel coupling potentials which are suppose to correct for missing reaction channels do not resolve this systematic discrepancy and one might recourse to the complex corrections factors in the future.
 
Extensive studies in SMEC \cite{Okolowicz2012,Okolowicz2013} have demonstrated that the low-energy coexistence of the cluster-like and shell-model-like configurations can be reconciled in the open quantum system formulation of shell model. The proximity of the branching point at threshold induces a collective mixing in shell-model-like states mediated by the aligned state that shares many features of the decay channel. This salient phenomenon in open quantum systems has been studied for the first time using the GSM-CC. The signature of a profound change of the near-threshold shell model wave function and of the direct manifestation of the continuum-coupling induced correlations is the presence of cluster states near their corresponding cluster emission thresholds. We have shown on few examples of the states in $^7$Li, $^7$Be how $^3$H, $^3$He, cluster correlations appear when the shell-model-like state approaches cluster decay threshold, and how these correlations faint again further away from the threshold. The appearance of a cluster is therefore associated with the collective response in other channels due to the unitarity. 
\\

\begin{acknowledgments}
Discussions with Witek Nazarewicz are gratefully acknowledged. This work has been supported by the National Natural Science Foundation of China under Grant Nos. 11975282; the Strategic Priority Research Program of Chinese Academy of Sciences under Grant No. XDB34000000; the State Key Laboratory of Nuclear Physics and Technology, Peking University under Grant No. NPT2020KFY13. 

\end{acknowledgments}

\bibliographystyle{apsrev4-1}

\bibliography{refs}

\end{document}